\documentclass[sigconf]{acmart}
\AtBeginDocument{%
  }

\usepackage{soul}
\usepackage{subcaption}
\usepackage[inline]{enumitem}
\usepackage{graphicx}
\usepackage{multirow}
\usepackage{tabularx}
\usepackage{array}
\usepackage{geometry}  
\usepackage{xcolor}

\newif\iftrackchanges
\trackchangesfalse  

\newcounter{hypothesis}
\renewcommand{\thehypothesis}{H\arabic{hypothesis}}
\newcommand{\hypothesis}[2]{%
    \refstepcounter{hypothesis}\label{#1}\textbf{\thehypothesis: #2}%
}

\copyrightyear{2025}
\acmYear{2025}
\setcopyright{cc}
\setcctype{by}
\acmConference[CHI '25]{CHI Conference on Human Factors in Computing Systems}{April 26-May 1, 2025}{Yokohama, Japan}
\acmBooktitle{CHI Conference on Human Factors in Computing Systems (CHI '25), April 26-May 1, 2025, Yokohama, Japan}\acmDOI{10.1145/3706598.3713751}
\acmISBN{979-8-4007-1394-1/25/04}

\begin{document}

\title[What Lies Beneath?]{What Lies Beneath? Exploring the Impact of Underlying AI Model Updates in AI-Infused Systems}

\author{Vikram Mohanty}
\orcid{0000-0001-6296-3134}
\affiliation{%
  \institution{\mbox{Human-Computer Interaction Institute}\\Carnegie Mellon University}
  \city{Pittsburgh}
  \state{PA}
  \country{USA}
}
\email{vikrammohanty@acm.org}
\authornote{These authors conducted substantial work on this project while at Virginia Tech.}

\author{Jude Lim}
\affiliation{%
  \institution{Independent Researcher}
  \city{Arlington}
  \state{VA}
  \country{USA}
}
\email{lkyoong428@gmail.com}
\authornotemark[1] 

\author{Kurt Luther}
\orcid{0000-0003-1809-626}
\affiliation{%
  \institution{Department of Computer Science \& \mbox{Center for Human-Computer Interaction}\\ Virginia Tech}
  \city{Alexandria}
  \state{VA}
  \country{USA}
}
\email{kluther@vt.edu}

\renewcommand{\shortauthors}{Mohanty et al.}

\begin{abstract}
AI models are constantly evolving, with new versions released frequently. Human-AI interaction guidelines encourage notifying users about changes in model capabilities, ideally supported by thorough benchmarking. However, as AI systems integrate into domain-specific workflows, exhaustive benchmarking can become impractical, often resulting in silent or minimally communicated updates. This raises critical questions: Can users notice these updates? What cues do they rely on to distinguish between models? How do such changes affect their behavior and task performance? We address these questions through two studies in the context of facial recognition for historical photo identification: an online experiment examining users’ ability to detect model updates, followed by a diary study exploring perceptions in a real-world deployment. Our findings highlight challenges in noticing AI model updates, their impact on downstream user behavior and performance, and how they lead users to develop divergent folk theories. Drawing on these insights, we discuss strategies for effectively communicating model updates in AI-infused systems.
\end{abstract}

\ccsdesc[100]{Information systems~Users and interactive retrieval}
\ccsdesc[300]{Computing methodologies~Computer vision tasks}
\ccsdesc[300]{Human-centered computing~Laboratory experiments}
\ccsdesc[500]{Human-centered computing~User studies}
\ccsdesc[500]{Human-centered computing~Empirical studies in HCI}

\keywords{AI Model Updates, Crowdsourcing, Facial Recognition, User Perception of AI Models, User Behavior, Historical Photo Identification, Folk Theories, Diary Study, Quantitative Methods}
  
\maketitle

\section{Introduction}

\begin{quote} In September 2023, Tesla rolled out a new beta version of its Full Self-Driving (FSD) software, promising advanced autonomous capabilities. Yet, for one driver, the update took an unexpected turn. While cruising on the highway, the car abruptly swerved toward a median at high speed—forcing the driver to intervene to avoid a crash. Unfortunately, it was not an isolated case; several other drivers reported similar issues, leading to widespread confusion about what had changed.~\cite{cao_2023} \end{quote}

AI model developers are pushing out new models and updates at a rapid pace, driven by growing user needs and the necessity to incorporate newer capabilities and address inaccuracies. For instance, Hugging Face hosts over 300,000 models~\cite{tousignant2023huggingface}, and in 2023 alone, 149 new foundation models were released~\cite{maslej2024aiindex}. These models power a wide range of AI-infused applications across domains such as healthcare, finance, social media, and autonomous vehicles~\cite{maslej2024aiindex}. 
However, with the ubiquity of AI-infused applications, many lay end-users may not fully understand AI model capabilities~\cite{pew2023ai,nader2024public}, let alone recognize or comprehend changes introduced by updates. This lack of awareness raises important questions: \textbf{Can users notice updates in AI models? What cues do they rely on to infer changes in underlying AI models? How do such updates influence their behavior, perceptions, and task performance?} 

Understanding how users perceive and adapt to these updates is critical, as updates often aim to improve performance but can unintentionally disrupt workflows, misalign expectations, or even lead to significant user dissatisfaction. Prior work has shown that model updates do not always improve human-AI team performance~\cite{bansal2019updates,chattopadhyay2017evaluating} and can lead to user dissatisfaction due to misaligned expectations or disruptions in user workflows~\cite{kocielnik2019will,bucher2012want}. Such unintended consequences can manifest in various ways, be it unforeseen accidents following updates to autonomous vehicle algorithms~\cite{cao_2023}, or distress among social media users due to alterations in recommendation algorithms~\cite{register2023attached}, or diminished volunteer engagement in citizen science projects upon the integration of automated systems~\cite{trouille2019citizen}. To avoid unexpected shifts in expectations,  human-AI interaction guidelines advocate for notifying users about model changes and capabilities~\cite{amershi2019guidelines}. 

Effectively communicating an AI model’s capabilities requires a deep understanding of what the model can and cannot do, which in turn demands comprehensive benchmarking and stress testing across diverse scenarios. However, as these models are deployed in workflows or domains that developers did not originally anticipate or design for, the complexity of thorough benchmarking increases~\cite{raji2021ai}. This makes comprehensive testing both impractical and costly, leading to silent updates or minimal communication or approximation about changes and capabilities, which leaves users unaware of significant model modifications. When users are unaware of these changes, they may interact with the system based on outdated assumptions, leading to potential errors, inefficiencies, or frustrations~\cite{bansal2019beyond,villareale2022want}.



To understand how users perceive and adapt to AI model updates in practical scenarios, we studied a facial recognition-based system used for historical person identification tasks (i.e., finding potential matches from a ranked list of search results for a given query photo). We conducted two complementary studies: 1) a controlled experiment simulating user interactions without explicit awareness of model updates, and 2) a real-world deployment where users could toggle between the old and new models.

For the first study, we ran an online experiment on Prolific to determine whether users could distinguish changes between facial recognition models across successive trials and what cues they relied on to do so. We also explored how the underlying model impacted their task performance, perceptions of accuracy and behavior, particularly when they were unaware of the model change. We complemented this with a second study in a real-world setting on Civil War Photo Sleuth (CWPS)~\cite{mohanty2019photo}, a widely-used platform for identifying historical photos through a facial recognition-based workflow. In April 2023, CWPS added a new facial recognition model alongside the existing one, allowing users to toggle between the two models. Through a two-week long diary study with 10 active users, we aimed to understand how they perceived the new model in contrast to the old one, and what factors influenced their model preferences.

Our findings provide key insights into users' abilities to detect model changes, the cues they rely on, and how these changes influence their task performance, perceptions of accuracy, behavior, and the folk theories and preferences they develop. Based on these findings, we discuss implications for deploying model updates in a human-AI collaborative environment.

\section{Related Work}

\subsection{User Frustration with Software Updates}

Software updates are ubiquitous in modern life, from operating system patches to app updates across devices. Most updates happen silently, with users often scheduling them for convenience, like overnight updates. However, some updates can have major, unforeseen consequences, such as the 2024 CrowdStrike incident that briefly disrupted global operations~\cite{enwiki:1244987724}, or stock exchange glitches caused by a system update~\cite{nyseglitch2023}. At the consumer level, prior work by Vaniea et al.~\cite{vaniea2016tales} shows that users go through a six-stage process when updating software, and unexpected changes, such as interface or functionality shifts, often lead to negative perceptions. Users frequently avoid updates when they anticipate disruptions to their workflows or loss of preferred features.

Though experts recommend auto-updates for security, many users disable them after experiencing issues like performance drops, preferring control over updates~\cite{mathur2017impact}. Haney et al.~\cite{haney2023user} found that smart home users frequently face confusion from unclear update notifications and compatibility issues, suggesting that clearer communication and more control, such as manual vs. automatic updates, would improve the experience. Mathur et al.~\cite{mathur2018quantifying} further emphasize that users avoid updates due to perceived costs (e.g., time, storage, risks like data loss), with unclear messaging exacerbating this reluctance. Morreale et al.~\cite{morreale2020my} further illustrate how software updates can drastically alter user experiences, as seen with Spotify's 2019 update, which removed features and disrupted user workflows. This update led to widespread frustration as users lost control over how they organized and accessed their music, highlighting how updates can impose new norms of use that reduce user agency. 

This resistance to updates is reflected in the findings from Rula et al.~\cite{rula2020s}, which showed that many users continue to run outdated software, avoiding updates due to concerns about potential disruptions or changes. This behavior further demonstrates how user apprehension can hinder the adoption of updates, leading to increased security vulnerabilities and reduced system performance. While these studies highlight software updates, AI model updates introduce new challenges that differ significantly from traditional software changes. Unlike software updates, which often involve visible changes to functionality or interfaces, AI model updates can occur silently, with changes in behavior or performance that are subtle or opaque to users. These updates, especially in black-box systems, may not provide clear cues that allow users to recognize when a change has occurred.

In our studies, we explore \textbf{whether users can notice changes in the underlying AI model}, and if so, \textbf{what cues they rely on}, and \textbf{how they characterize the differences between the models}. Understanding whether users can distinguish such updates is critical, as the inability to do so might lead to similar downstream effects on user trust, performance, or decision-making.

\subsection{User Perceptions of Dynamic AI Systems}

Advances in AI research have opened new avenues for human-AI collaboration, leveraging the complementary strengths of humans and AI systems to drive progress across diverse domains such as creativity, manufacturing, and journalism~\cite{wilson2018collaborative,liu20233dall,broussard2019artificial}. However, this collaboration is far from static or fully understood. Human users and AI agents do not interact in fixed ways; instead, their behaviors and interactions evolve dynamically, influenced by factors such as context of use, task complexity, system outputs, and the continually evolving, yet often uncertain, nature of AI technologies~\cite{yang2020re,amershi2019guidelines,horvitz1999principles,shavit2023practices}.


AI-assisted decision-making --- an example of \emph{human-AI teaming}, where humans make decisions based on AI-provided suggestions and recommendations~\cite{bansal2019beyond,berretta2023defining,lai2023towards} --- illustrates this interplay. While such collaboration has led to significant advancements, it also introduces challenges. In these workflows, pitfalls such as automation bias, amplification of incorrect results, and the perpetuation of existing biases can undermine trust and effectiveness, highlighting the need for a more nuanced understanding of human-AI dynamics~\cite{willever2014family,jones2023people,kim2024m}.

Prior research strongly advocates for a nuanced understanding of both human and AI capabilities to foster more effective collaboration~\cite{bansal2019beyond,wang2021towards,cai2019hello,he2022walking,liao2021human}. A substantial body of literature has explored the factors shaping user perception of AI, including algorithm speed~\cite{park2019slow,teevan2013slow,gnewuch2018faster}, the number of results generated~\cite{oulasvirta2009more}, model explanations~\cite{kim2023help,wang2021explanations,wang2023watch,vasconcelos2023explanations}, trust calibration interventions~\cite{ma2023should,buccinca2021trust}, group behavior~\cite{chiang2023two}, and the system’s ability to communicate its expertise effectively~\cite{zhang2022you}. Research also shows that providing users with some degree of control can enhance their perception of and trust in the system~\cite{kocielnik2019will}.

As the market experiences a rapid influx of new AI models~\cite{cnbc_alibaba_ai_models_2024,wiggers_mistral_ai_models_2024,nunez_nvidia_ai_model_2024,reuters_ibm_ai_models_2024,reuters_amazon_ai_models_2024}, this frequent turnover, while often beneficial, presents unique challenges. One of the key questions is \textbf{how users perceive and adapt to changes in AI systems, particularly when updates to the underlying models are not clearly communicated}. Prior work has explored how users may perceive shifts in model explanations after updates, even when the decisions remain unchanged~\cite{wang2023watch}. These shifts can affect users' trust, particularly if the new explanations diverge significantly from previous ones. Users may experience cognitive dissonance when their established mental models are violated, leading to reduced trust and satisfaction with the system. 

However, there remains a notable gap in fully understanding user perceptions, particularly concerning updates in black-box models deployed in real-world scenarios. Prior research indicates that the introduction of new models is not always welcomed by users. In some cases, it can lead to decreased engagement and increased distress due to the unpredictability of the system's behavior~\cite{register2023attached,trouille2019citizen}. This highlights the need for a more comprehensive understanding of how users react to major model upgrades, especially in systems where the internal workings are opaque and explanations cannot be easily communicated.

In this work, we explore how users perceive AI model updates both in a controlled environment and a real-world deployment, focusing on the cues they rely on and how these updates influence their performance and behavior. Examining these elements in practical contexts of AI model updates can help in validating human-AI interaction guidelines, particularly those related to effectively communicating model changes~\cite{amershi2019guidelines}.

\subsection{Civil War Photo Sleuth and Historical Person Identification}

Civil War Photo Sleuth (CWPS)\footnote{https://www.civilwarphotosleuth.com/} is an online platform designed to help users identify unknown individuals in historical photos through a facial recognition-based pipeline~\cite{mohanty2019photo}. By presenting a list of visually similar candidates, CWPS supports users in performing side-by-side facial similarity comparisons, a fairly generic visual inspection task~\cite{mohanty2023photo}. The platform’s established user base is already accustomed to interacting with AI-driven face recognition, making it an ideal setting for studying user perceptions of AI model updates. In the context of the historical person identification task, users rely on both the AI’s recommendations (i.e., search pool of facially similar-looking candidates) and their own subjective judgment to make decisions (i.e., find potential matches from the search pool), offering a unique opportunity to observe how different models influence decision-making. With over 50,000 photos in its database, CWPS provides a rich platform for investigating how users respond to varying model outputs. 

For the first study, we developed a web platform that closely mirrored CWPS's search interface and used their API to retrieve facially similar candidates for running experiments on Prolific. In the second study, we recruited 10 active CWPS users to compare the outputs of two distinct facial recognition models. We will explore the design, methodology, and findings of these studies in detail in the following sections.

\section{Study 1: Distinguishing AI model updates without explicit communication}

In this study, we investigate how updates in underlying AI models influence user perceptions, performance, and behaviors in environments where changes occur without explicit communication. To explore this, we conducted an online experiment where participants completed multiple trials of a historical person identification task. In each trial, they were asked to find matches for a given query photo by interacting with the results of a randomly assigned black-box facial recognition model.

Participants were not informed about which model they were using or if the model had changed between trials. Through this study, we examine \textbf{1) whether users can accurately detect changes in the underlying model}, \textbf{2) what kind of cues they rely on to make this distinction}, and \textbf{3) how their perceptions, performance, and behaviors shift when the model changes}.

\subsection{Hypotheses} \label{sec:hypotheses}


We build on prior work and the key differences between the two models (see Section \ref{sec:facial-rec-models}) to propose hypotheses guiding our investigation into whether users can detect changes in an underlying AI model, the cues they rely on, and how these updates influence their behavior and performance. Prior work has shown that users can detect subtle shifts in system behavior, primarily when provided with model explanations in non-black-box, controlled settings~\cite{wang2023watch}. Real-world cases, such as user discussions on Reddit about perceived changes in GitHub Copilot’s code suggestions and backlash to Replika updates that altered conversational behaviors~\cite{reddit_github_copilot,cole_replika_2023}, indicate that users may notice changes in AI behavior even without explicit communication. Extending these findings, we hypothesize \textbf{\hypothesis{hyp:H1} Users will be able to distinguish a change in the model even though they are interacting with black-box systems.}

We investigate these questions within the context of a historical person identification task where users examine AI-retrieved search results (i.e., a ranked list of visually similar candidates) and combine these with their own judgment to identify potential matches. In such retrieval tasks, like web search engines, observable cues like latency and the number of results displayed shape how users perceive and interact with the system, potentially serving as key indicators for distinguishing between models. Prior work demonstrates that users are sensitive to latency, with even small delays being noticed and impacting satisfaction~\cite{brutlag2009speed,teevan2013slow}. Similarly, the number of results shapes user perception: too many options can overwhelm users, while fewer results increase focus but risk neglecting potentially relevant items~\cite{oulasvirta2009more,kelly2015many}. Building on these insights, we hypothesize:
\textbf{\hypothesis{hyp:H2} Users will rely on observable system characteristics, such as the number of search results and latency, to distinguish between models.}

In addition to examining cues for distinguishing between models, we also investigate how users perceive the accuracy of the underlying face recognition models. Based on cognitive science theories of similarity, users are likely to perceive the model that retrieves candidates that are more similar-looking as more accurate because they can detect subtle improvements in output quality by focusing on high-diagnostic and directly comparable features~\cite{markman1996commonalities,tversky1977features}. Similarly, prior research in search engines suggests that fewer, more precise results --- a trait of the newer face recognition model we tested (see Section \ref{sec:facial-rec-models}) --- can increase user trust in the system by reducing cognitive load and emphasizing relevance~\cite{oulasvirta2009more}. These findings inform \textbf{\hypothesis{hyp:H3} When comparing two models (older vs. newer), users will perceive the newer model (deemed more accurate by developers) as more accurate, even if they are not explicitly aware of which model they are using or informed about model changes.}

Updating an AI model to improve its standalone accuracy raises the question of whether these improvements translate to better human-AI team performance. While higher model accuracy intuitively reduces errors and enhances outcomes~\cite{towler2023diverse}, prior research highlights that this may not always hold true, as mismatches between AI improvements and human workflows can hinder collaboration~\cite{bansal2019updates,chattopadhyay2017evaluating}. Despite these complexities, we hypothesize \textbf{\hypothesis{hyp:H4} Human-AI team performance will improve with the more accurate model.}

Model updates in AI systems can introduce changes that subtly alter user interactions, prompting adjustments in behavior. Prior research on software updates shows that unexpected changes to functionality or outputs can lead users to modify their workflows, engagement levels, or decision-making processes~\cite{morreale2020my,mathur2018quantifying}. Similarly, changes in recommender algorithms, as observed in social media platforms, have been shown to influence user behavior by shifting how they explore or engage with content based on perceived output quality,  relevance, or volume~\cite{neimanlab2019,karizat2021algorithmic}. In the context of a retrieval task, such shifts might manifest as differences in the number of results users check, the time spent on tasks, or their decision-making strategies. Building on these insights, we hypothesize: \textbf{\hypothesis{hyp:H5} AI model updates will affect user behavior in terms of time spent on tasks, the number of results they check, and the decisions they make.}

Moreover, when users believe a model has been updated, even if this belief is not explicitly confirmed, it can lead to behavioral changes driven by cognitive dissonance. According to Cognitive Dissonance Theory~\cite{festinger1957theory}, users experiencing a mismatch between their expectations and observed system behavior may adjust their actions to reduce discomfort, altering their decision-making and interaction patterns. To explore this further, we hypothesize: \textbf{\hypothesis{hyp:H6} When users believe that a model has switched, it will lead to changes in their behavior.}

\subsection{Experiment Setup}

To explore these hypotheses, we designed an online survey experiment where participants interacted with a facial recognition-based web application (see Figure~\ref{fig:exploring-search-results}), modeled after the search results page on \textit{CWPS}~\cite{mohanty2019photo}. In each trial, participants were shown a query photo and tasked with identifying potential matches (i.e., other photos of the same person) from a ranked list of search results retrieved by one of two distinct facial recognition models (see Section~\ref{sec:facial-rec-models} for details). The assignment of models was randomized behind the scenes, and participants were not informed which model was being used or if the model changed between trials. Participants visually inspected the results to determine whether any of the retrieved images matched the person in the query photo.

Each participant completed 8 trials, with each trial featuring a different query photo and its corresponding search results (i.e., potential candidates). This randomization ensured that, while participants were unaware of the underlying model, we could systematically explore their ability to detect changes between the models and the cues they relied on. This design enabled us to examine shifts in user perceptions, performance, and behaviors across trials.

\subsection{Facial Recognition Models}\label{sec:facial-rec-models}

Similar to CWPS~\cite{mohanty2019photo}, we used facial recognition models from Microsoft Azure for this study: the oldest available model, \textit{recognition\_01} (released in 2017), and the most recent model, \textit{recognition\_04} (released in 2021). For clarity, we will refer to these as the \textbf{old model} and \textbf{new model}, respectively throughout this paper. According to the developer website~\cite{urban_2023}, the new model is supposed to be the most accurate available, designed to handle challenging cases like faces with masks or facial hair. However, its effectiveness in specialized domains, such as historical photo identification, remains unclear. Previous benchmarking of the old model on CWPS showed that while effective at retrieving correct matches, it also retrieved a large number of search results (mostly false positives), increasing the likelihood of incorrect matches~\cite{mohanty2020photo}.

We set the confidence threshold for retrieving results at 0.50 based on prior work~\cite{mohanty2020photo}, which demonstrated its utility in avoiding false negatives (i.e., missing the correct match) while ensuring a manageable set of retrieved results for the older model. However, given the newer model's substantially improved precision and recall (Appendix~\ref{appendix:benchmarking}), this threshold produced significantly fewer results: an average of 20 compared to 378 with the older model. While this large discrepancy in result count might suggest a system-level design difference, it actually reflects the inherent capabilities of the models under identical configurations. To preserve these native configurations and better simulate real-world usage scenarios, we conducted our study without artificially constraining result counts. However, we acknowledge this as a potential limitation in Section~\ref{sec:limitations}.

Our benchmarking study (see details in Appendix ~\ref{appendix:benchmarking}) on the dataset we used for the study (see Section~\ref{sec:dataset}) revealed notable differences between the two models: \textbf{the new model consistently outperformed the old model} in terms of \textbf{average precision (i.e., ranking correct matches higher); 0.79 (new) vs. 0.41 (old)} and \textbf{recall (i.e., finding more correct matches); 0.76 (new) vs. 0.54 (old)}. 
Additionally, while the old model retrieved a large number of results, many were false positives, making it relatively less precise in this domain. In contrast, the new model produced fewer results overall but retrieved more correct matches with higher confidence scores.

These differences in key metric s--- such as the number of results, the ranking of more similar-looking candidates, and improved handling of diversity --- make the old model and new model compelling candidates for this study. However, while these measurable differences exist, an important question remains: \textbf{can users actually perceive these differences in practice?} Since participants were unaware of which model was in use during each trial, the study explores whether users can rely on visible cues, such as fewer but more accurate results and better-ranked matches, to distinguish between the two models. 

\subsubsection{Latency in retrieving results} Prior work has demonstrated that latency, i.e., the time taken by a system to retrieve result, can significantly shape user perceptions of algorithmic systems, where even slight delays reduce engagement and satisfaction in search and chatbot contexts, but in some cases, slower responses may foster reflection or higher perceived quality~\cite{park2019slow,teevan2013slow,brutlag2009speed,gnewuch2018faster}. In this study, we introduced an artificial delay to explore \textbf{whether users consider latency when trying to distinguish between model changes and how it may influence their perception of model accuracy}. Participants were assigned to one of four conditions: \textbf{(i)} Old Model retrieves results faster than New Model, \textbf{(ii)} New Model retrieves results faster than Old Model, \textbf{(iii)} both models retrieve results with random latency, or \textbf{(iv)} both models retrieve results with identical latency. For slower conditions, we introduced a 12-second delay, while standard conditions took 3 to 5 seconds. 

\subsection{Interface}

We designed the web application to guide participants through the person identification tasks across eight consecutive trials. In each trial, participants were shown a ranked list of search results for the query photo, retrieved by one of the two models described in Section~\ref{sec:facial-rec-models} (the \textbf{old model} or the \textbf{new model}). The specific model for each trial was randomly assigned, and participants were blinded to the model’s identity (no old/new labels were shown) or whether the model had changed between trials. This ensured that participants’ assessment of the models were based solely on their interaction with the model outputs, without any influence from model labels or associations. The interface supported five key components:

\begin{enumerate}
    
    \item \textbf{Exploring Search Results}: In each trial, participants were presented with a ranked list of potential candidates retrieved by a facial recognition model for a given query photo. The list was ordered by the model’s confidence score, which indicated how likely each candidate resembled the query photo (see Figure~\ref{fig:exploring-search-results}). Participants visually inspected this list of candidates to find possible matches for the query photo.

    \begin{figure}[h]
        \centering
        \includegraphics[width=\linewidth]{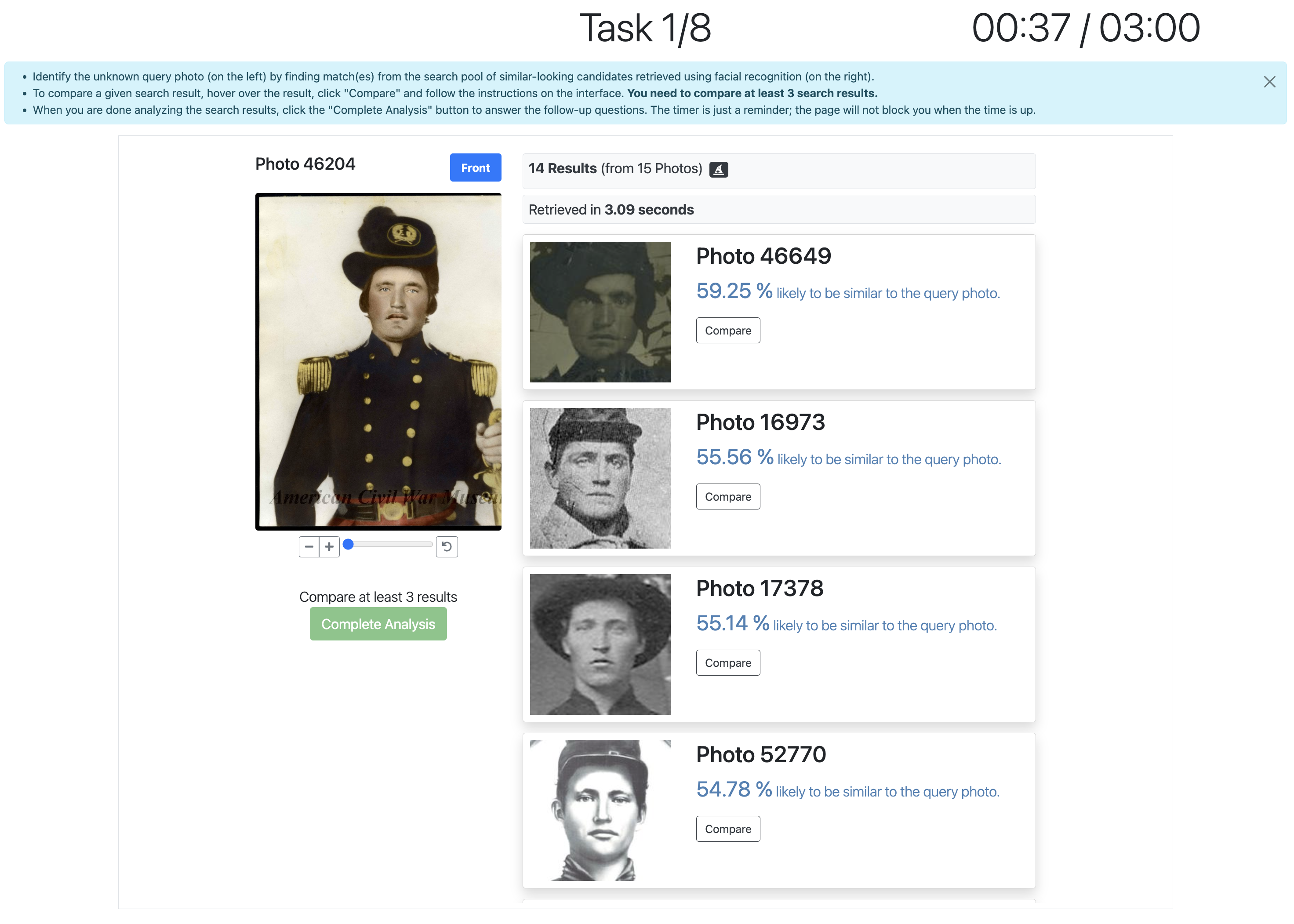}
        \caption{Search interface for examining potential matches retrieved by the facial recognition model. The search results (right) are ranked according to the model’s confidence score, which ranges from 0\% (not at all similar-looking) to 100\% (highly similar-looking). Only candidates with confidence scores above 50\% were retrieved, following the CWPS threshold. Each search result includes a unique photo ID (from CWPS), thumbnail of the face, and the model's confidence score, quantifying its similarity to the query photo (on the left). Participants were required to compare at least three search results before completing the analysis.}
        \Description{A search interface displaying a historical query photo on the left and a list of potential matches on the right, each accompanied by a confidence score. The list is ranked by confidence scores, ranging from 54.78\% to 59.25\%, indicating how similar each candidate is to the query photo. The interface requires participants to compare at least three results before completing the analysis.}
        \label{fig:exploring-search-results}        
    \end{figure}
    
    \item \textbf{Comparing Results}: Participants can perform a close inspection on any search result by opening an interface that allows them to do a side-by-side comparison of the query photo and the selected search result (similar to \cite{mohanty2023photo}). This comparison interface allows users to then vote on whether the two photos showed a \textit{facial match} (i.e., same person, different view), a \textit{replica} (i.e., same person, same view), \textit{different people}, or if they were \textit{not sure} (see Figure~\ref{fig:compare-interface}).

    \begin{figure}[h]
        \centering
        \includegraphics[width=\linewidth]{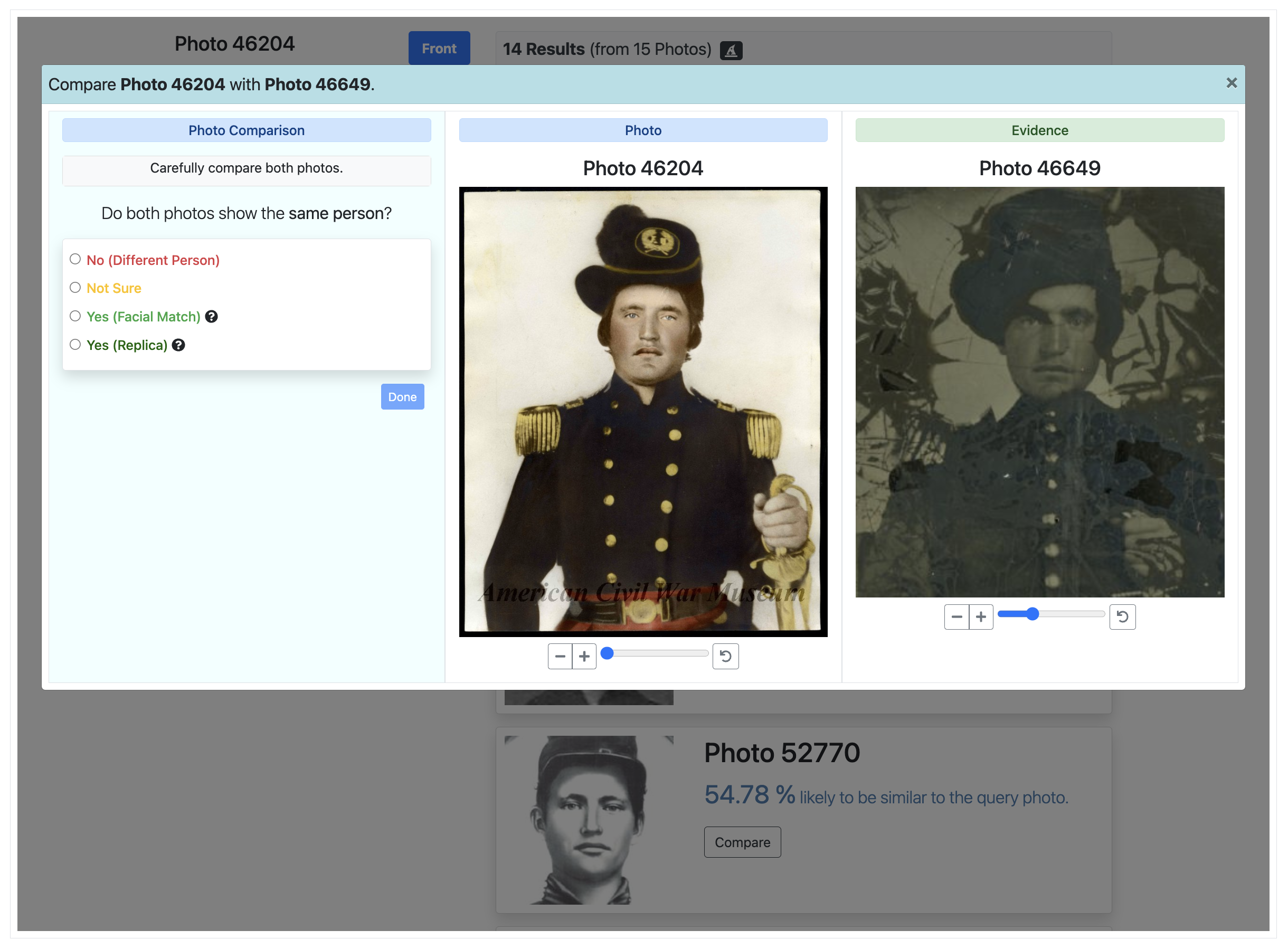}
        \caption{Side-by-side comparison interface for close inspection of a search result (right) with the query photo (left). Participants can carefully compare the two images and vote on one of four options: \textit{facial match} (same person, different view), \textit{replica} (same person, same view), \textit{different people}, or \textit{not sure}.}
        \Description{Side-by-side comparison interface showing a historical query photo on the left and a search result on the right. Participants can choose from four options after comparing the two photos: facial match, replica, different people, or not sure.}
        \label{fig:compare-interface}
    \end{figure}    
    
    \item \textbf{Rating Model Accuracy}: At the end of each trial (i.e., after participants finished comparing the search results), they rated the accuracy of the facial recognition model on a slider ranging from -100 (very inaccurate) to 100 (very accurate), with an optional text box to explain their rating (see Figure~\ref{fig:model-rating}). We chose this wide range to give participants more flexibility to express shifts in perceived accuracy across trials, without attributing precise or fixed meaning to any specific point on the scale.

    \begin{figure}
        \centering
        \includegraphics[width=\linewidth]{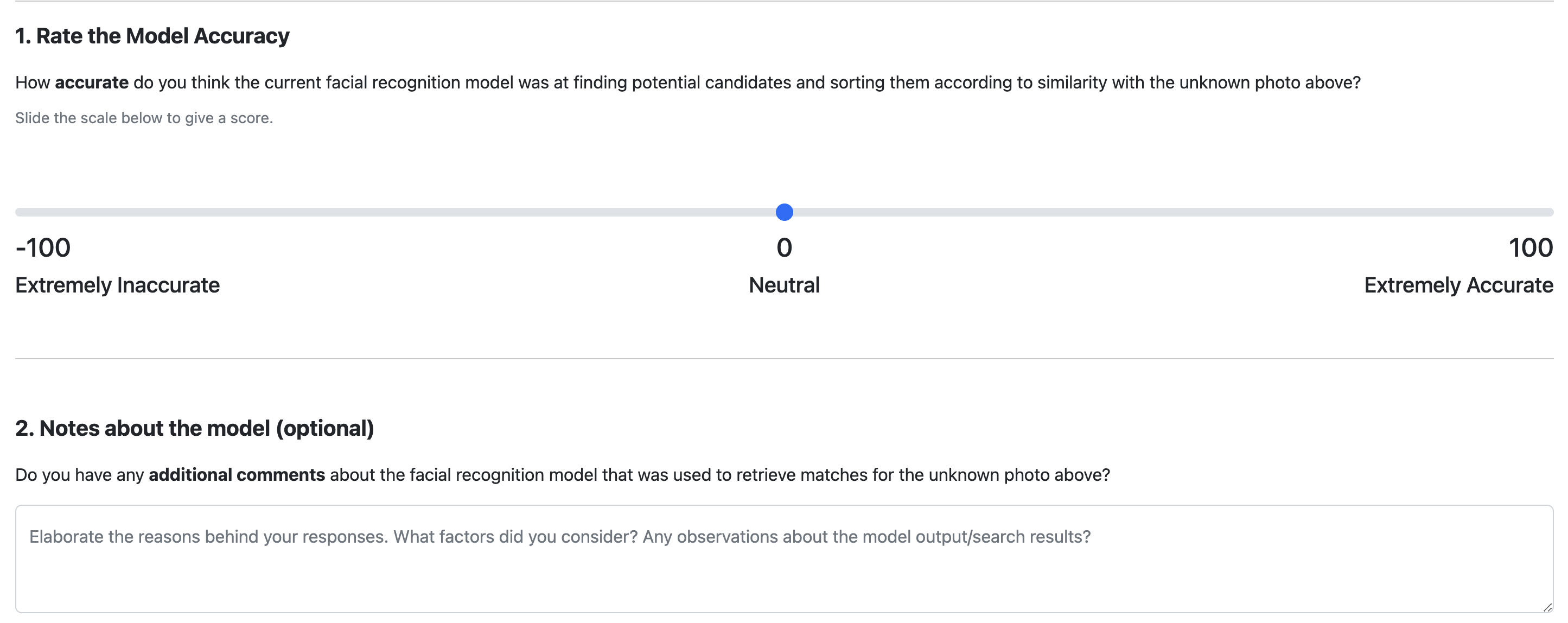}
        \caption{Model rating interface where participants rated the accuracy of the facial recognition model used in the current trial. The slider ranged from -100 (extremely inaccurate) to 100 (extremely accurate). An optional text box was provided for participants to explain their rating, allowing them to elaborate on the factors that influenced their decision.}
        \Description{A model rating interface showing a slider for participants to rate the accuracy of the facial recognition model. The scale ranges from -100 (extremely inaccurate) to 100 (extremely accurate), with an optional text box below to provide additional comments on the rating.}
        \label{fig:model-rating}
    \end{figure}
    
    \item \textbf{Comparing Models Across Trials}: After rating the current model, participants compared the model used in the current trial with the one from the previous trial (see Figure~\ref{fig:model-comparison}). The interface displayed relevant trial-specific details such as the query photo, model latency (time taken to retrieve results), the number of search results, whether the participant found a match, and their perceived accuracy. Participants then rated the similarity or difference between the models on a four-point Likert scale: \textit{very different, somewhat different, somewhat similar}, or \textit{very similar}. 

    The labeling of trials followed a consistent process based on participants’ responses. The model in the first trial was always labeled as \textbf{Model A} by default. For each subsequent trial, participants compared the model in the current trial with the one from the previous trial and indicated whether they believed it was the \textit{same} model or a \textit{different} model:

    \begin{itemize}
        \item If participants indicated the model was \textit{different}, the current trial was assigned the opposite label from the previous trial (e.g., if the previous trial was labeled \textbf{Model A}, the current trial was labeled \textbf{Model B}, and vice versa).

        \item If participants indicated the model was \textit{the same}, the current trial retained the same label as the previous trial (e.g., if the previous trial was \textbf{Model A}, the current trial also remained \textbf{Model A}).
    \end{itemize}

    This process created a sequence of labels (Model A or Model B) that reflected participants’ perceptions of whether the model changed or remained consistent across trials. It is important to note that these labels (Model A and Model B) represent user-perceived models and do not correspond to the actual underlying model (\textbf{old} or \textbf{new}) used in each trial, which was randomized independently.
    

    \begin{figure}
        \centering
        \includegraphics[width=\linewidth]{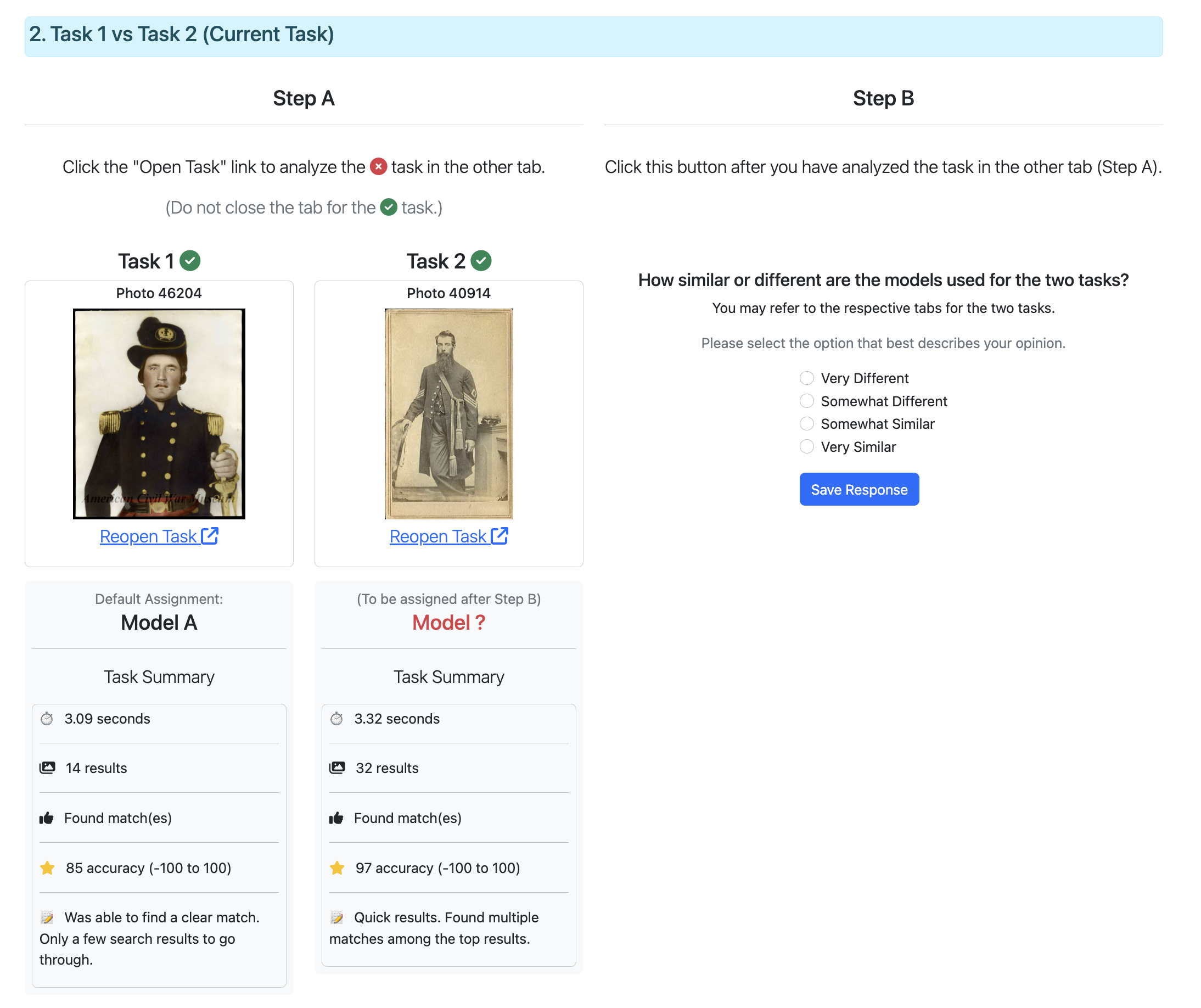}
        \caption{Interface for comparing the model used in the current trial with the model from the previous trial. The interface displays trial-specific details, including the query photo, model latency, number of search results, whether the participant found a match, and the participant’s perceived accuracy. Participants then rate the similarity or difference between the models on a four-point Likert scale: \textit{very different, somewhat different, somewhat similar}, or \textit{very similar}.}
        \Description{Interface for comparing the models used in two consecutive trials. It shows the query photos, task summaries, model latency, number of search results, and perceived accuracy for each trial. Participants are asked to rate how similar or different the models used in the trials are.}
        \label{fig:model-comparison}
    \end{figure}

    \item \textbf{Selecting a Preferred Model}: After all 8 trials were completed, participants were shown a summary of their trial categorizations (see Figure~\ref{fig:model-preference}).
    This step in the workflow was designed to help participants organize their thoughts about the trials they categorized as \textbf{Model A} or \textbf{Model B}. The interface displayed key trial-specific information, such as task duration, the number of search results, whether a match was found, and the participant's accuracy rating for each trial. Based on this summary, participants selected their preferred model for future photo identification tasks and provided a justification for their choice.

    \begin{figure}
        \centering
        \includegraphics[width=\linewidth]{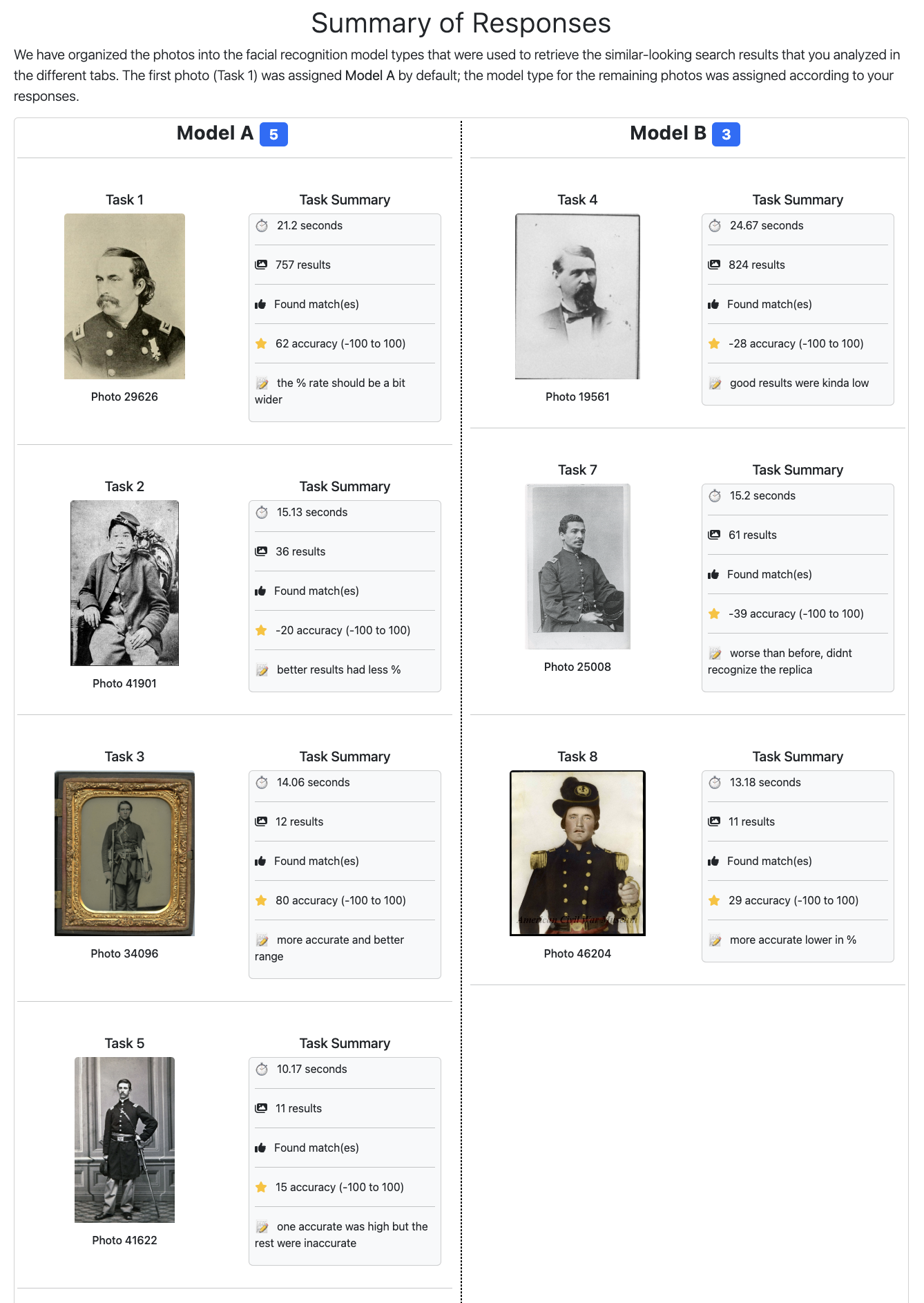}
        \caption{Summary of all trials, categorized into \textbf{Model A} and \textbf{Model B} as perceived by the user. The first trial was automatically assigned to \textbf{Model A}, while the remaining trials were categorized based on the participant’s model comparison responses. The interface displays trial-specific information such as task duration, number of search results, and the participant’s accuracy rating. After viewing the summary, participants were asked to select their preferred model for future photo identification tasks and provide justification for their choice.}
        \Description{A summary interface showing all trials categorized into Model A and Model B based on user perception. The first trial is assigned to Model A by default, while the remaining trials are categorized according to the participant's comparison responses. Each trial displays details such as task duration, number of search results, and the participant's accuracy rating.}
        \label{fig:model-preference}
    \end{figure}

\end{enumerate}

To distinguish the user-perceived model labels from the actual underlying models (the old and new models, which were unknown to participants), we will refer to Model A and Model B as \textit{\textbf{Model A-UP}} and \textit{\textbf{Model B-UP}} (user-perceived) for clarity.

\subsection{Dataset}\label{sec:dataset}

Our dataset consists of 20 diverse Civil War portraits sourced from the CWPS database, similar to prior studies using this dataset~\cite{mohanty2020photo,mohanty2019second}. The pool includes images of Union and Confederate soldiers of different ranks, predominantly white males, as well as African American soldiers. We also added photos of two white females, an Asian, and a Hispanic soldier. Some photos have known positive matches in the database, while others do not, allowing us to capture varying model performance across a range of inputs. This selection reflects the over-representation of white males in the CWPS dataset, which stems from historical biases~\cite{kusuma2022civil}. Additionally, facial recognition models are known to have racial and gender biases, often performing less accurately on non-white and female faces~\cite{buolamwini2018gender,raji2019actionable}. As a result, we anticipated that these biases might influence participants' perceptions of the models' performance, especially in cases involving women and people of color. 

For each participant, we randomly selected 7 photos from the 20-photo pool. These 7 photos were distributed across 8 trials, with one photo intentionally repeated using two different models in separate trials. This design facilitated a direct comparison between models for the same query photo, allowing participants to assess potential differences in model performance with a familiar visual reference. The search results, or potential candidates presented to participants in each trial, were retrieved by the facial recognition models from the CWPS database, ensuring that all photos came from the same search pool of relevant candidates.

\subsection{Participants}

We recruited a total of 252 participants globally (58.7\% men and 41.3\% women; mean age = 29.8 years, SD = 9.7 years) through Prolific~\cite{prolific_2024}. Participants were all at least 18 years of age, fluent in English, and hailed from multiple countries. The countries with the most participants were South Africa (57 participants), Poland (46 participants), and the United Kingdom (39 participants). All participants completed an IRB-approved consent form before beginning the task. The study was approved by the university’s IRB.

\subsection{Measurement}

We collected several interaction and behavioral data points from 252 participants, each completing 8 trials, which resulted in a total of \textbf{2016 search sessions} (see Table~\ref{tab:data-collected}). In total, participants made 10646 comparison decisions across trials and completed 1764 pairwise model comparisons (7 per user). The \textbf{new model} was (randomly) assigned to 1015 search sessions, while the \textbf{old model} was assigned to 1001 sessions. The sample size of 252 participants was chosen to ensure sufficient data for robust statistical analysis while remaining feasible within the constraints of the study design.

\begin{table*}[h]
\centering
\renewcommand{\arraystretch}{1.2}  
\newcolumntype{Y}{>{\raggedright\arraybackslash}X}
\begin{tabularx}{\textwidth}{|p{5cm}|Y|}
\hline
\textbf{Measurement} &
  \textbf{Description} \\ \hline
\textbf{Time spent on search tasks} &
  Overall time spent on the search results page and individual comparison times. \\ \hline
\textbf{Comparison decisions} &
  Decisions made during side-by-side comparisons, with participants choosing between: \textit{facial match}, \textit{replica}, \textit{different people}, or \textit{not sure}. \\ \hline
\textbf{Perceived accuracy of the model} &
  Slider ratings from -100 (very inaccurate) to 100 (very accurate) after each trial and an optional note justifying their decision. \\ \hline
\textbf{Model comparisons} &
  Participants rated how similar or different the models were between two consecutive trials using a four-point Likert scale: \textit{very different}, \textit{somewhat different}, \textit{somewhat similar}, or \textit{very similar}. \\ \hline
\textbf{Final model preference} &
  At the conclusion of the study, participants selected their preferred model between Model A-UP and Model B-UP for future tasks and provided a justification. \\ \hline
\end{tabularx}
\vspace{0.25cm}
\caption{Table summarizing the key measurements collected during the study, including participant behavior, decisions, and model evaluations over the 8 trials.}
\label{tab:data-collected}
\end{table*}

In order to understand the signals participants used to distinguish between models, we also calculated the following "difference" metrics between consecutive trials for each participant, as participants compared these factors side-by-side during model comparisons (see Figure 4):

\begin{itemize}
    \item \textbf{Difference in perceived accuracy:} The change in participants' perceived accuracy ratings (from -100 to 100) between two consecutive trials.
    \item \textbf{Difference in the number of search results retrieved: } The difference in the number of results returned by the models in each trial, a key observable factor during model comparisons.
    \item \textbf{Difference in response latency:} The time difference in the models’ response time (latency), as varying response times were introduced to explore their effect on participant perceptions.
\end{itemize}

To evaluate participants' identification performance with the different models, we calculated the following metrics based on comparisons with the ground truth:

\begin{itemize}
    \item \textbf{Precision}: The proportion of correct identifications (true positives) out of all results flagged as a positive match by participants (i.e., classified as either a facial match or a replica). This measures how accurately participants identified correct matches.
    \item \textbf{Recall}: The proportion of correct identifications (true positives) out of all actual positive matches available in the search results. This measures how well participants were able to find all the correct matches in the results retrieved by the model.
    \item \textbf{False Positive Rate (FPR)}: For trials where no correct matches were present in the results, FPR was calculated as the proportion of incorrect matches (false positives) classified by participants as either a facial match or a replica. This helps us assess how often participants mistakenly identified incorrect results as positive matches.
\end{itemize}

The selected metrics were chosen to comprehensively evaluate the impact of model changes on both user performance and perceptions. Precision, recall, and false positive rate (FPR) are standard metrics for assessing the effectiveness of facial recognition systems, providing objective measures of identification accuracy, completeness, and error rates. Perceived accuracy and difference metrics were included to capture users’ subjective evaluations and the observable cues they relied on when comparing models. Specifically, we focused on the number of search results and response latency, as these are typical observable cues in retrieval tasks and directly influence user perception and behavior~\cite{oulasvirta2009more,brutlag2009speed}. Additionally, we recorded participants’ comparison decisions and the time taken to make these decisions, as these metrics are particularly relevant in the context of this study, where users are comparing search results to identify matches. Together, these metrics provide a balanced view of both objective system performance and subjective user experience, enabling us to address our research questions comprehensively.

\subsection{Analysis}

To answer our research questions, we employed the following statistical methods: 

\begin{itemize}

    \item \textbf{Paired t-tests:} Used to compare precision, recall, and false positive rates (FPR) between the models to assess improvements in identification performance. We also applied t-tests to examine participants' ability to detect model changes and to compare perceived accuracy between the models. This was used for the following findings: Sections~\ref{sec:finding_h1},~\ref{sec:findings_h3}, and~\ref{sec:hat_performance}.

    \item \textbf{Logistic mixed-effects models (GLMM):} Used to predict binary outcomes such as participants’ ability to detect a model change (a binary outcome) and to analyze their final model preference. Key predictors included differences in perceived accuracy, latency, and search results between consecutive trials. GLMM was also applied to predict comparison decisions (e.g., positive vs. negative vs. uncertain decisions), using factors such as the underlying model type and time taken for individual comparisons. The fixed-effect coefficients (\(\beta\)) represent the magnitude and direction of the relationship between predictors and the log odds of the binary outcome. Exponentiating \( \beta \) provides the corresponding odds ratio, which indicates the multiplicative change in odds for a one-unit change in the predictor. For instance, a one-unit increase in perceived accuracy difference (\(\beta = -0.89630\)) reduces the odds of detecting a model switch by approximately 59.2\%, corresponding to an odds ratio of 0.408 (\(e^{-0.89630} \approx 0.408\)). We implemented GLMM using the $glmer$ function in R~\cite{bates2014fitting}. This was used for the following findings: Sections~\ref{sec:findings_h2} and ~\ref{sec:model_preference}.

    \item \textbf{Linear mixed models (LMM):} Used to predict continuous outcomes, such as participants' perceived accuracy of the models and various behavioral metrics (e.g., time spent making decisions, number of results checked). Key predictors included participants' interactions with the models, such as the number of search results, the types of comparison decisions (e.g., facial match or replica), and the underlying model type. The fixed-effect coefficients (\(\beta\)) represent the magnitude and direction of the relationship between predictors and the outcome (e.g., perceived accuracy). For instance, each additional facial match response (\(\beta = 23.54\)) increases perceived accuracy by 23.54 units, while scrolling through more results (\(\beta = -4.75\)) decreases it by 4.75 units. We implemented LMM using the $lmer$ function in R~\cite{bates2014fitting}. This was used for the following findings: Sections~\ref{sec:findings_perceived_accuracy_factors},~\ref{sec:findings_h5}, and ~\ref{sec:findings_h6}.

\end{itemize}

Mixed-effects models (both logistic and linear) were used to account for the nested structure of the data (multiple trials per participant), with random intercepts included to control for individual variability.
\subsection{Findings}

\subsubsection{\textbf{Participants struggled to distinguish the underlying facial recognition model used between two consecutive trials.}}\label{sec:finding_h1}

Out of 1764 pairwise model comparisons across 252 participants, the overall accuracy in detecting whether models had changed was \textbf{48.87\%}, \textbf{close to random guessing}. When the model changed, participants identified the change \textbf{56.98\%} of the time, but when the model remained the same, they accurately detected this only \textbf{38.38\%}. This difference in detection rates was statistically significant, with participants being better at identifying model changes than noticing when the model remained the same (\(t(1665.5) = -7.91\), \(p < 0.001\)). These findings contradict \ref{hyp:H1}, as participants generally struggled to distinguish changes between models, especially when no change occurred.

Participants’ comments revealed a notable inconsistency in how they interpreted perceived changes or continuity in the model’s behavior. Some participants misinterpreted changes in the model’s behavior as inherent variability or adaptation within a single model, rather than recognizing distinct systems. For instance, one participant remarked: \emph{"Same AI? They got the two same matches,"} after observing similar outputs (e.g., common matches) across different trials, suggesting they assumed consistency even when the model had changed. Conversely, other participants attributed perceived shifts in behavior to a different model, even when the underlying model remained the same. For example, one participant noted: \emph{"This model seems to be matching photos with consideration for ethnic features"}, indicating that they perceived these shifts as stemming from a new system.


\subsubsection{\textbf{Participants primarily relied on their perceived accuracy of the models rather than observable characteristics like latency or result count.}}\label{sec:findings_h2}

When comparing models across successive trials, participants were shown the latency, number of search results, and their own perceived accuracy for both trials. Differences in \textbf{perceived accuracy played a key role} in participants’ ability to distinguish between models. Specifically, for every one-unit increase in perceived accuracy difference, the likelihood of identifying the models as different increased substantially ($\beta = -0.89630$, $z = -12.122$, $p < 0.001$). This indicates that \textbf{as participants sensed greater differences in accuracy, they were much more likely to detect model changes.}

On the other hand, differences in observable characteristics like \textbf{latency} and \textbf{result count} did not significantly affect participants’ ability to distinguish between models (latency: $\beta = 0.06054$, $z = 1.005$, $p = 0.315$; result count: $\beta = 0.03753$, $z = 0.635$, $p = 0.525$). These findings directly contradict \ref{hyp:H2}, as participants primarily relied on their subjective assessment of model performance rather than observable characteristics such as latency or the number of search results when distinguishing models between trials.

\subsubsection{\textbf{Participants perceived the new model as more accurate than the old model, despite not knowing which model they were interacting with.}}\label{sec:findings_h3}

Across 2016 trials, participants rated the perceived accuracy of the models on a scale (see Figure~\ref{fig:model-rating}) from -100 (highly inaccurate) to 100 (highly accurate). Participants were tasked with determining whether the model they used in each trial was the same as or different from the previous trial, but they were never informed whether they were interacting with the old model or the new model. As shown in Finding~\ref{sec:finding_h1}, participants struggled to distinguish whether the model changed between trials, meaning any potential conscious mapping to the actual underlying model (old or new) was effectively lost. These ratings were based solely on their interaction with the outputs in each trial, without any knowledge of the underlying model's identity. During analysis, we retrospectively mapped trials to the actual models (old or new) used, enabling us to compare perceived accuracy across the two models.

\begin{figure}[h]
    \centering
    \includegraphics[width=\linewidth]{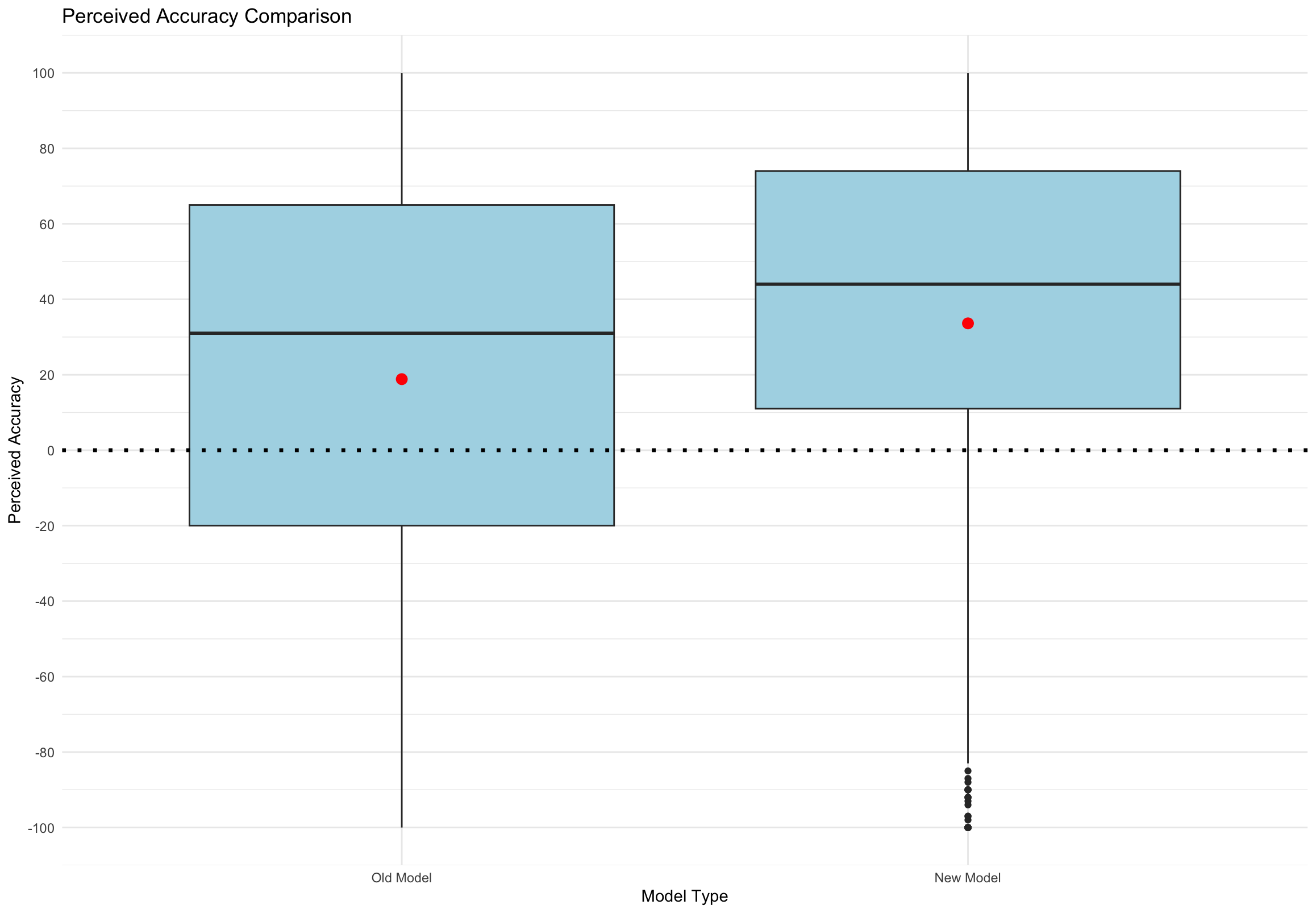}
    \caption{Participants perceived the \textbf{new model }to be more accurate compared to the \textbf{old model} despite moderate absolute accuracy ratings. The mean is denoted by the red dots in the boxplots.}
    \Description{Boxplot comparing perceived accuracy between two models: old model and new model. The new model's boxplot shows higher perceived accuracy, with the mean indicated by a red dot. The new model also has a few outliers below -80, but overall participants rated it significantly higher than the old model.}
    \label{fig:perceived-accuracy-model-comparison}
\end{figure}

Despite this, participants consistently rated the \textbf{new model} as significantly more accurate (\textbf{mean accuracy = 33.62}) compared to the \textbf{old model} (\textbf{mean accuracy = 18.32}; \(t(1981.5) = -6.10\), \(p < 0.001\)), thus supporting \ref{hyp:H3}. It is also worth noting that \textbf{neither model was perceived as highly accurate}, with both models receiving moderate ratings on the scale (see Figure~\ref{fig:perceived-accuracy-model-comparison}).

\subsubsection{\textbf{User perceptions of AI model accuracy were largely shaped by their interactions with model outputs and engagement behaviors.}}\label{sec:findings_perceived_accuracy_factors}

\begin{figure*}[h]
    \centering
    \includegraphics[width=\linewidth]{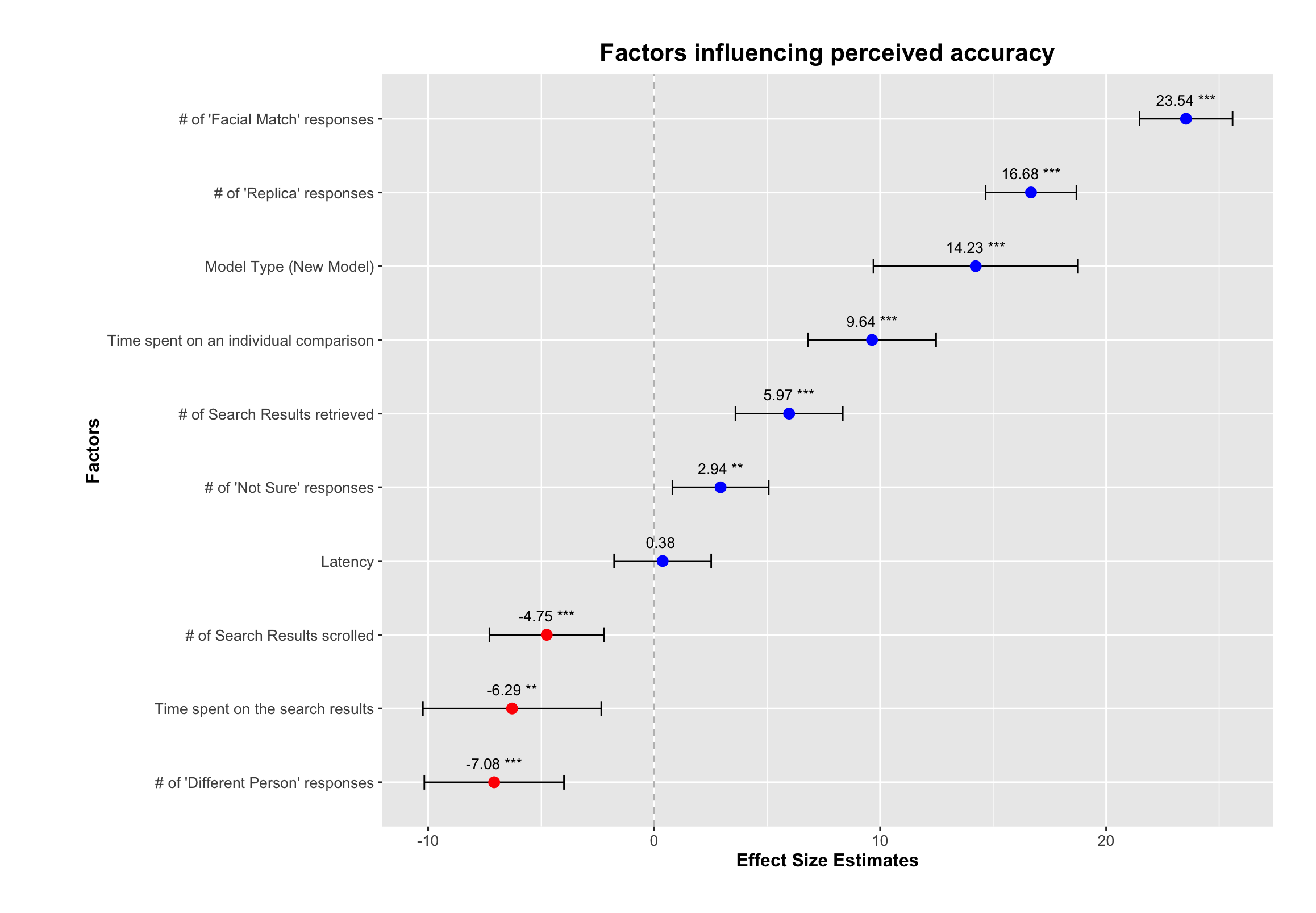}
    \caption{Factors influencing perceived accuracy. The plot shows effect size estimates (in units of perceived accuracy) for various factors. Positive factors such as \textbf{facial match responses} (\(\beta = 23.54\), meaning that each additional facial match response increases perceived accuracy by 23.54 points) and \textbf{replica responses} (\(\beta = 16.68\)) had the largest impact on increasing perceived accuracy. In contrast, \textbf{different person responses} decreased accuracy (\(\beta = -7.08\), meaning each additional response of this type reduces perceived accuracy by 7.08 points). \textbf{Scrolling through more search results} (\(\beta = -4.75\)) and \textbf{time spent on the search page} (\(\beta = -6.29\)) negatively impacted perceptions, while the \textbf{number of search results retrieved} had a small positive effect (\(\beta = 5.97\)). Stars next to effect sizes denote statistical significance, with \textbf{***} indicating \(p < 0.001\), \textbf{**} indicating \(p < 0.01\), and \textbf{*} indicating \(p < 0.05\).}
    \Description{A dot-and-whisker plot showing the effect size estimates for various factors influencing perceived accuracy. Positive effects (in blue) include \textbf{facial match responses} (\(\beta = 23.54\)) and \textbf{replica responses} (\(\beta = 16.68\)), while negative effects (in red) include \textbf{different person responses} (\(\beta = -7.08\)) and \textbf{scrolling through more search results} (\(\beta = -4.75\)). \textbf{Latency} shows no significant effect (\(\beta = 0.38\)). The chart has stars next to the effect size to indicate statistical significance.}
    \label{fig:perceived-accuracy-factors}
\end{figure*}

We observed that when participants made \textbf{positive comparisons}, it influenced perceived accuracy in a positive way, with more \textbf{facial match responses} (\(\beta = 23.54, p < 0.001\)) and \textbf{replica responses} (\(\beta = 16.68, p < 0.001\)) leading to higher perceived accuracy (see Figure~\ref{fig:perceived-accuracy-factors}). In contrast, \textbf{negative comparisons}, such as identifying a \textbf{Different Person}, significantly reduced perceived accuracy (\(\beta = -7.08, p < 0.001\)). 

Participant comments reflected these findings, highlighting that personal experiences while comparing results strongly shaped their perceptions. One participant remarked, \emph{"It appeared to retrieve a match as the top result but some of the other results looked nothing like the original person"}, reinforcing how accuracy perceptions were influenced by direct comparison outcomes. Another noted, \emph{"Facial hair seems to be a hindrance for facial recognition. Although some seemed slightly similar, none were correct"}, pointing to specific features impacting perceived accuracy.

We also found that \textbf{user behavior} played an important role in shaping perceptions. Spending more time on individual comparisons positively influenced perceived accuracy (\(\beta = 9.64, p < 0.001\)). However, \textbf{scrolling through more search results} and spending additional time on the overall search page negatively affected perceived accuracy (\(\beta = -4.75, p < 0.001\); \(\beta = -6.29, p < 0.01\), respectively). Participants' comments also pointed to concerns about diversity and representation in the model’s outputs, which influenced their perception of its accuracy. One participant stated, \emph{"The selection of data it was trained on lacked validity because model output had more males than females in comparison to the female query photo"}. Another noted, \emph{"The race of the person on the photo and the race of the most similar matches the model suggested were different"}, pointing to racial mismatches, which diminished their confidence in the model.
    
Finally, when participants interacted with the \textbf{new model}, they were more likely to rate the model as having higher perceived accuracy (\(\beta = 14.23, p < 0.001\)), even though they were unaware of which underlying model they were using. This is in line with \ref{hyp:H3}. Interestingly, \textbf{latency} did not have a significant impact on perceived accuracy (\(p = 0.73\)), while a higher \textbf{number of search results} had a small but significant positive effect on perceived accuracy (\(\beta = 5.97, p < 0.001\)).

\subsubsection{\textbf{User preferences were significantly driven by perceived model accuracy}} \label{sec:model_preference}

Our analysis showed that \textbf{perceived accuracy} was a significant predictor of model preference.  For every unit increase in the perceived accuracy of \textbf{Model A-UP}, the odds of preferring Model A-UP \textbf{decreased by 3.4\%} ($\beta = -0.03398$, $z = -4.94$, $p < 0.001$). Conversely, for every unit increase in the perceived accuracy of \textbf{Model B-UP}, the odds of preferring Model B-UP \textbf{increased by 3.8\%}  ($\beta = 0.03799$, $z = 5.69$, $p < 0.001$). Other factors, such as the number of trials assigned to each model or whether a match was found, did not significantly influence model preference (all $p > 0.05$).

\subsubsection{\textbf{Human-AI team performance showed minimal improvement in recall with the newer model, while precision and false positive rates remained similar.}}\label{sec:hat_performance}

When comparing participants' decisions to ground truth across all 2016 search sessions, we observed only a marginal improvement in \textbf{recall} with the \textbf{new model} (62.64\%) compared to the \textbf{old model} (56.05\%), as confirmed by a two-sample Welch t-test ($t(1274.1) = -2.99$, $p = 0.0028$). However, there was no significant difference in \textbf{precision} or \textbf{false positive rates} between the models (see Table~\ref{tab:humanaiteamperformance}). These findings suggest that \ref{hyp:H4} was not fully supported, as improvements in Human-AI team performance with the more accurate model were minimal.

\begin{table}[h]
\small
\centering
\renewcommand{\arraystretch}{1.5} %
\begin{tabular}{l|cc|c}
\multirow{2}{*}{\textbf{Models}} &
  \multicolumn{2}{c|}{\textbf{\begin{tabular}[c]{@{}c@{}}At least 1 positive match\\ in the search results\end{tabular}}} &
  \textbf{\begin{tabular}[c]{@{}c@{}}No positive match \\ in the search results\end{tabular}} \\ \cline{2-4} 
                                & \textbf{Precision} & \textbf{Recall} & \multicolumn{1}{l}{\textbf{False Positive Rate}} \\ \hline
\multicolumn{1}{c|}{\textbf{Old Model}} & 80.59\%            &\textbf{ 56.05\%}         & 25.40\%                                          \\
\multicolumn{1}{c|}{\textbf{New Model}} & 81.10\%            & \textbf{62.64\% }        & 27.5\%                                          
\end{tabular}
\vspace{0.25cm}
\caption{User performance with the old and new facial recognition model across 2016 search sessions. The metrics used are mean precision, recall, and false positive rates.}
\label{tab:humanaiteamperformance}
\end{table}

\subsubsection{\textbf{The underlying model significantly influenced user behavior in terms of time spent and the kind of decisions made}}\label{sec:findings_h5}

While interacting with the outputs of the newer model, participants made \textbf{1 fewer comparison} on average ($\beta = -1.00$, $t = -2.86$, $p < 0.01$), spent \textbf{16.7 seconds less} per session ($\beta = -16.74$, $t = -5.03$, $p = 5.46 \times 10^{-7}$), and scrolled through \textbf{11.6 fewer results} ($\beta = -11.60$, $t = -13.38$, $p < 2 \times 10^{-16}$) compared to the older model.

Our findings further revealed that the underlying model significantly impacted both the \textit{decisions} participants made and the \textit{time} they spent engaging with the system. When interacting with the outputs of the new model, participants were \textbf{46\% more likely to make positive decisions} (i.e., classifying an image as a match or replica) compared to the old model ($\beta = 0.38$, $z = 7.76$, $p < 0.001$), and \textbf{30.7\% less likely to make negative decisions} ($\beta = -0.37$, $z = -8.11$, $p < 0.001$). The new model did not significantly affect uncertain decisions ($\beta = 0.08$, $z = 1.43$, $p = 0.15$).

Participants also took \textbf{0.61 seconds longer} per comparison with the new model ($\beta = 0.61$, $t = 3.72$, $p = 2 \times 10^{-4}$). Interestingly, the time spent per comparison was linked to decision type: \textbf{longer times were associated with an increased likelihood of positive} ($\beta = 0.046$, $z = 15.24$, $p < 2 \times 10^{-16}$) and \textbf{uncertain} decisions ($\beta = 0.040$, $z = 13.30$, $p < 2 \times 10^{-16}$), while \textbf{shorter times were linked to negative decisions} ($\beta = -0.099$, $z = -24.62$, $p < 2 \times 10^{-16}$). These findings support ~\ref{hyp:H5}, as the underlying model significantly influenced user behavior.

\subsubsection{\textbf{Perceived model changes had negligible impact on user behavior}}\label{sec:findings_h6}

Contrary to ~\ref{hyp:H6}, perceived model changes had a limited effect on user behavior. Participants who believed the model had switched did not significantly alter the number of comparisons made ($\beta = 0.24$, $t = 0.57$, $p = 0.57$), the time spent per session ($\beta = -0.06$, $t = -0.015$, $p = 0.99$), or the mean time spent per comparison ($\beta = -0.27$, $t = -1.04$, $p = 0.30$). However, participants did scroll through \textbf{2 more results} on average when they perceived a model switch ($\beta = 2.04$, $t = 1.98$, $p = 0.048$), suggesting slight exploratory behavior.

\subsection{Summary of Findings}

Despite significant performance differences between the two models, participants in the online experiment were \textbf{unable to accurately distinguish between them}, contradicting~\ref{hyp:H1}. This resulted in \textbf{no measurable improvement in performance}, even when interacting with the newer, more accurate model, and thus refuting~\ref{hyp:H4}. In order to assess the models, participants \textbf{relied heavily on their perceptions of accuracy}, formed through personal interactions with the system, rather than using objective metrics such as latency or result count. The newer model did, however, lead to \textbf{some changes in behavior}, with participants spending more time deliberating on positive or uncertain matches, supporting~\ref{hyp:H5}.

\section{Study 2: Comparing User Perceptions of Old and New AI Models in a Real-World Deployment}

Study 1 revealed that when model updates were \textbf{fully invisible}, users struggled to detect changes and did not fully capitalize on the more accurate model's potential, leading to only \textbf{marginal improvements in human-AI team performance}. In this study, we explore whether \textbf{explicitly informing} users of a new model would lead to more effective use of its outputs and how their \textbf{preferences and perceptions} would differ when directly comparing two models. 

We investigated how users of \textbf{Civil War Photo Sleuth (CWPS})---a platform where the \textbf{older facial recognition model had been in use}---perceived and compared a newly introduced AI model. These users, already familiar with the older model, were informed that a "new model" was available, but were given no specific details about its characteristics or improvements. Unlike Study 1, where model updates were hidden from participants, this study allowed users to explicitly \textbf{toggle between the old and new models} (same as Study 1) in a real-world environment, giving them the agency to explore differences between the models on their own terms.

Our goal was to uncover not only user preferences but also the \textbf{factors influencing these preferences} and\textbf{ the folk theories} users developed about the models' performance in a practical, real-world setting.

\subsection{Study Setup}

\begin{figure}[h]
    \centering
    \begin{subfigure}[b]{0.45\textwidth}
        \centering
        \includegraphics[width=\textwidth]{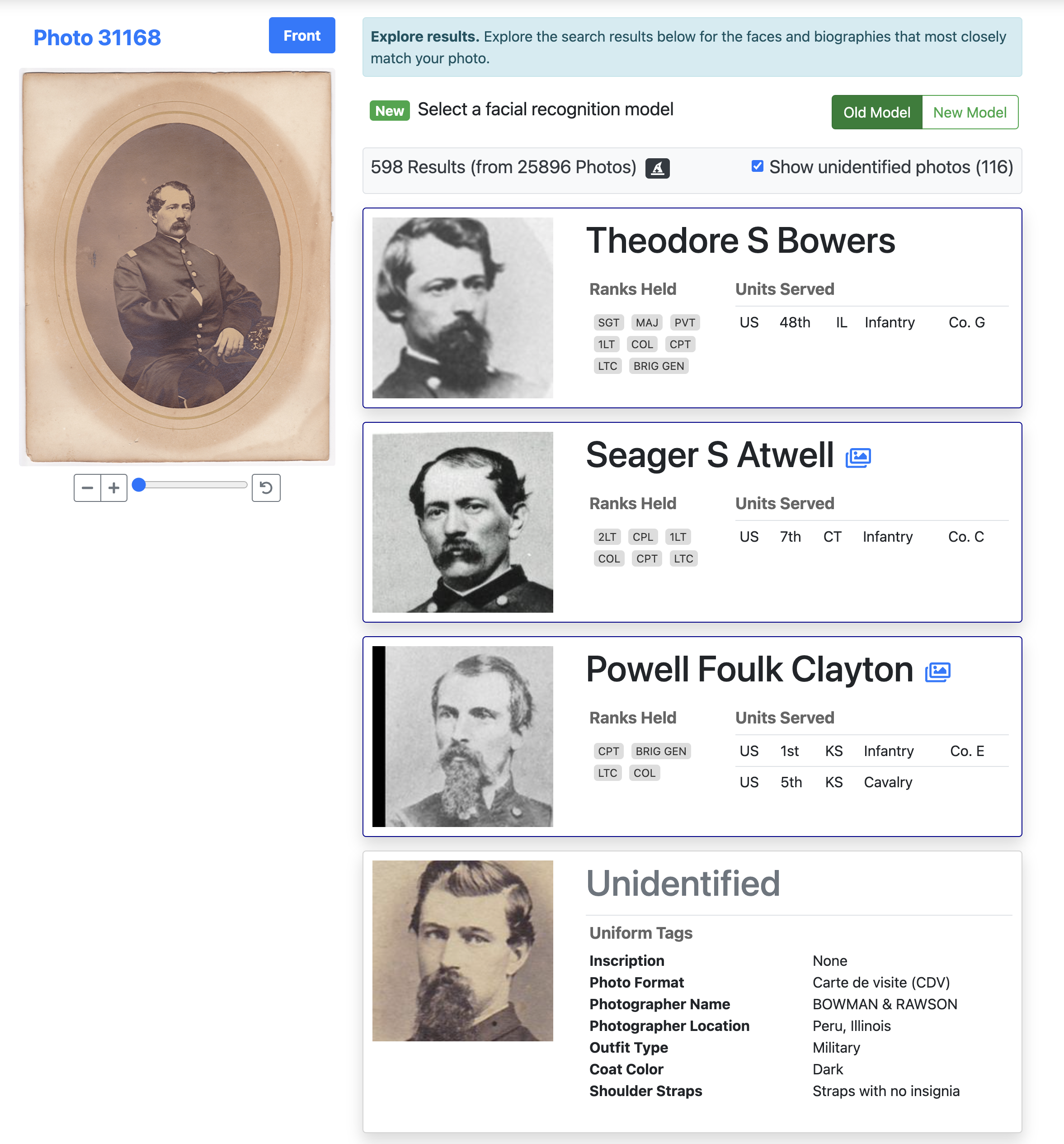}
        \caption{Results from the old model: 598 results retrieved.}
        \Description{The search interface of Civil War Photo Sleuth showing results from the old model. The interface displays 598 retrieved results with images of historical figures, including Theodore S. Bowers, Seager S. Atwell, Powell F. Clayton, and an unidentified person.}
        \label{fig:oldmodel}
    \end{subfigure}
    \hfill
    \begin{subfigure}[b]{0.45\textwidth}
        \centering
        \includegraphics[width=\textwidth]{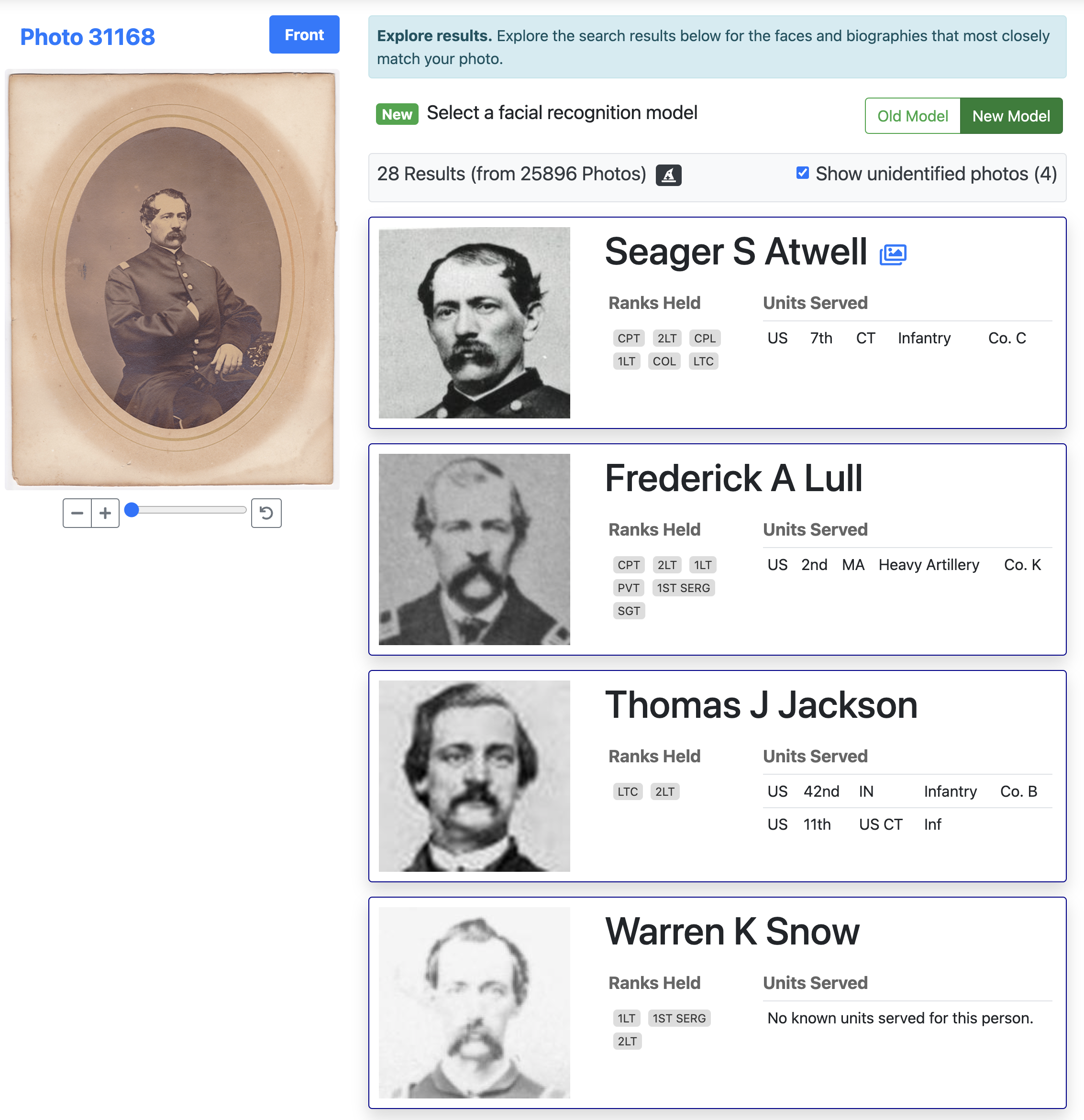}
        \caption{Results from the new model: 28 results retrieved.}
        \Description{The search interface of Civil War Photo Sleuth showing results from the new model. The interface displays 28 retrieved results with images of historical figures, including Seager S. Atwell, Frederick A. Lull, Thomas J. Jackson, and Warren K. Snow.}
        \label{fig:newmodel}
    \end{subfigure}
    \caption{Comparison of search results retrieved by the old model (a) and the new model (b). The old model retrieved a significantly larger number of results (598) compared to the new model (28). In addition to fewer results, the new model presents different people, suggesting improvements in the relevance of the results retrieved.}
    \label{fig:sidebyside}
\end{figure}

In April 2023, the CWPS team integrated the most recent facial recognition model (i.e., the \textbf{new model} from Study 1) alongside the existing older model (see Section~\ref{sec:facial-rec-models} for model details). Users could seamlessly \textbf{switch between the outputs of both models via a toggle button} added to the search interface, allowing them to continue their regular photo investigation tasks (see Figure~\ref{fig:sidebyside}).

We recruited 10 CWPS users by posting an advertisement on social media to solicit interest and collaborating with the CWPS team to filter active users who had engaged with the platform in the past six months. The group comprised nine males and one female, with an average age of 46 (min = 19, max = 62). Each participant expressed a certain degree of familiarity with the platform's facial recognition search feature. Participants investigated the identities of 10 photos of their choice, either by uploading new images or selecting from existing ones on the site.

Over two weeks, they were instructed to use both models and document their reflections in a diary (see template in supplemental material), including which model they found more useful and why, along with a usefulness rating from -100 (\textit{not at all useful}) to 100 (\textit{highly useful}). A mid-point reflection was collected after five photo investigations, followed by a final reflection at the conclusion of the study to assess their overall preferences and perceptions of both models. This study was approved by the university’s IRB.

\subsection{Analysis}

We reviewed and categorized participants' diary entries, focusing specifically on responses provided in designated answer fields within the diary template. These fields captured participants' thoughts on the old and new models, their perceived strengths and weaknesses, and their reflections on similarities and differences between the models. 

One author conducted an open coding process to identify recurring concepts and observations within the answer fields. Through this process, we identified 24 unique codes, such as "Speed," "Photo Quality," "Result Accuracy," "Facial Features," "Unrelated Features," "Mixed Results," and "Prefer Old/New/Both". To ensure the robustness and consistency of the analysis, two authors collaboratively reviewed these codes and grouped them into seven overarching themes: \textit{"Preference," "Perceived Accuracy," "Result Counts," "Other Performance Factors," "Misc. Visual Factors," "Unrelated to Model," and "Behavior"}. This iterative process involved deliberation and consensus to ensure that the themes accurately reflected the diverse dimensions of participants’ reflections.

Finally, we categorized participants' responses according to these themes, and the findings are presented in the subsequent sections of the paper.

\subsection{Findings}

\subsubsection{\textbf{The new model was generally preferred, but not unanimously.}}

\begin{figure}[h]
    \centering
    \begin{subfigure}[b]{0.45\textwidth}
        \centering
        \includegraphics[width=\textwidth]{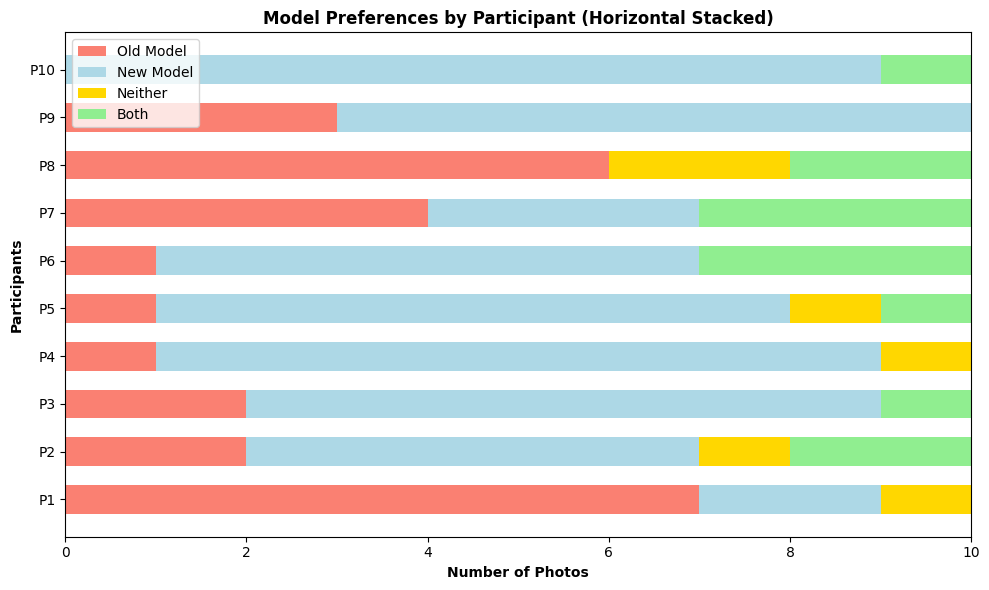}
        \caption{Model Preferences of Users.}
        \Description{The stacked horizontal bar chart illustrates the preferences of 10 participants (P1 to P10) regarding the use of old and new AI models, as well as options for using neither or both in tandem. Each bar represents a participant and is divided into sections representing the number of times they chose each option. The data indicates a diverse range of preferences across participants. Participants P4, P5, and P9 showed a strong preference for the new model, with P10 exclusively preferring it. P8 notably leaned towards the old model, while P1 demonstrated a balanced preference with a slight lean towards the old model. Participants P2, P3, P6, and P7 exhibited mixed preferences, with some even opting for using both models or neither at various points. Overall, the chart portrays a nuanced picture of user preferences, indicating that different individuals have varied approaches and levels of comfort with the old and new AI models.}
        \label{fig:modelpreferences}
    \end{subfigure}
    \hfill
    \begin{subfigure}[b]{0.45\textwidth}
        \centering
        \includegraphics[width=\textwidth]{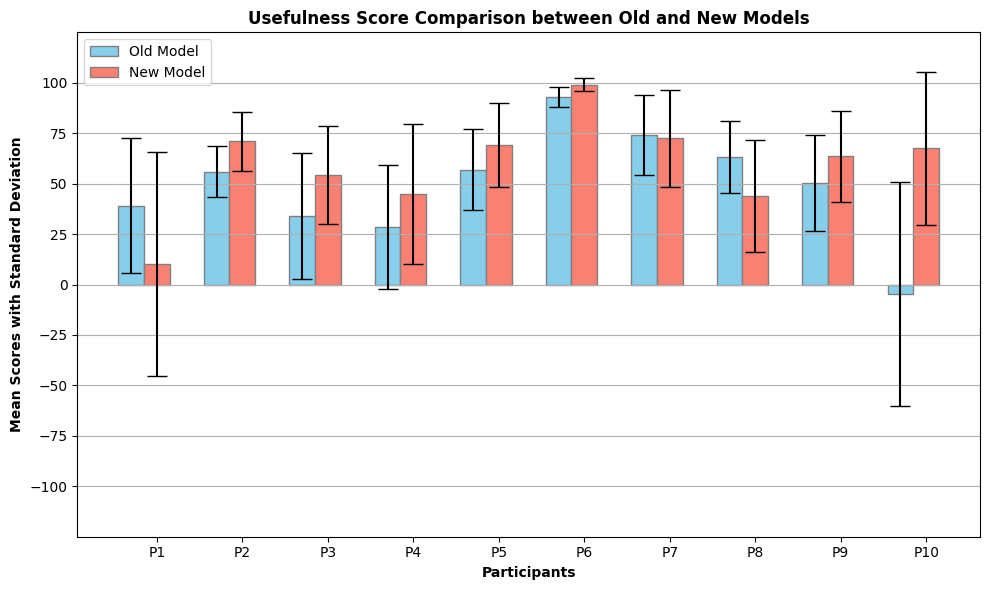}
        \caption{Usefulness Scores for Old and New Models.}
        \Description{The bar chart compares the mean usefulness scores between the old and new models as rated by each participant (P1 to P10). The chart displays two sets of bars for each participant: one representing the old model (in sky blue) and the other representing the new model (in salmon). Each bar signifies the mean score, while the error bars indicate the standard deviation, showcasing the variability in scores given by each participant. The participants are listed along the x-axis, and the mean scores, ranging from -100 to 100, are plotted along the y-axis, with grid lines at intervals of 25 units.}
        \label{fig:usefulnessscores}
    \end{subfigure}
    \caption{(a) Model Preferences of Users and (b) Usefulness Scores for Old and New Models in the Diary Study.}
    \label{fig:combined}
\end{figure}

While most participants favored the new model, citing its accuracy and relevance in matching images (7 out of 10 participants), the older model remained useful in certain cases. As shown in Figure \ref{fig:modelpreferences}, participants \textbf{chose the new model for 54 photos} and \textbf{the old model for 27}. In the words of \textbf{P10}: \textit{"The New Model is a clear winner in my book. I tried to challenge it using images that I knew contained the same
person but with different angles and conditions and it almost always passed. The potential for matching faces in outdoor
group images to studio portraits is tremendous here, and I think this will prove to be an invaluable tool."}

Some users did not notice substantial differences between the models. As P1 observed, \textit{"...I noticed little difference between the two. Both were fast, both produced similar results, as in the same possible people."} This was reflected in comparable scores for usefulness, as seen in Figure \ref{fig:usefulnessscores}. Moreover, there were 19 instances where participants opted for neither model exclusively, instead choosing both. \textbf{P7} suggested, \textit{"Using both models in tandem is the way to go rather than comparing one directly to the other."} This sentiment was echoed by \textbf{P2}, who \textit{preferred toggling between models to explore different perspectives}.

\subsubsection{\textbf{Cognitive load from sifting through large numbers of search results shaped user preferences.}}

Participants had mixed reactions to the number of search results, which significantly influenced their model preferences. Some users valued the \textbf{larger result set} from the old model for offering more choices and the \textbf{potential for serendipitous discoveries}. As P1 noted, \textit{"There were more of a revolving door of options [with the old model], whereas the new model only gave me five."} P8 echoed this, appreciating the exploratory aspect: \textit{"The fact that the image could possibly be buried way down the results page in the older model makes searching much more enjoyable."}

However, many participants found the \textbf{larger result set overwhelming}, citing the \textbf{increased cognitive load }when sifting through numerous results. P3 explained, \textit{"After scrolling through the first 50, you kinda give up on looking for any more close matches."} P6 preferred the \textbf{new model for its streamlined results}, stating, \textit{"It produces much fewer results, so I am not scrolling through an infinite number of possible matches with the old model."} Similarly, P10 found the old model overwhelming, remarking, \textit{"The Old Model yielded 574 results (vs. just 17 from the New), so it was a bit overwhelming."}

Overall, participants who favored \textbf{efficiency and lower cognitive load gravitated towards the new model}, while those who valued \textbf{more options and exploratory searches preferred the old model}.

\subsubsection{\textbf{Perceived accuracy shaped user preferences differently across models}}

Participants’ preferences were often shaped by their perceptions of each model’s accuracy, though this varied on a case-by-case basis. For instance, \textbf{P3} favored the new model for its accuracy in capturing facial features, noting, \textit{"Hairlines, eyes, ears, facial all seem to match better."} On the other hand, \textbf{P8} preferred the older model for its broader image characteristics, observing, \textit{"Seems that with the old model, the resulting images had characteristics closer to my image: chin structure, hairstyle, eye colors."}

Some participants observed that \textbf{both models were adept at finding matches}, though their performance varied slightly in terms of accuracy. \textbf{P5} shared an instance where both models successfully identified an alternate view of the same soldier, stating, \textit{"Both models were able to find an alternate view of the same soldier. Even though this one is not inscribed, I can say without a doubt that it’s the same soldier as another image of Andrew Johnson. Results were similar in that the different view of Johnson showed up as the top result in both the new and old model."} However, P5 gave a slight edge to the new model, noting that it showed \textit{"more facial similarity."} This sentiment was also observed by \textbf{P6} and \textbf{P9} in similar cases.

Participants also \textbf{shared instances where one model clearly outperformed the other}. \textbf{P4} observed that the old model found a match the new model missed: \textit{"That face turned up as \#2 in the Old Model, whereas it didn't even show up in the New Model."} Conversely, \textbf{P6} shared a case where the new model succeeded in identifying a known subject that the old model missed: \textit{"The new model correctly identified him with the very first subject."}

\textbf{Ranking accuracy} also played a role in how participants rated the models. \textbf{P4} noted that, while both models retrieved a similar result, the new model ranked the match higher, stating, \textit{"The New Model ranked the best matched photo highest, whereas the Old Model had it further down the queue."} In contrast, \textbf{P5} found the old model to be more accurate in certain cases.

\subsubsection{\textbf{The trade-off between quantity and quality influenced user preferences for each model}}

Participants often experienced a trade-off between the number of search results and their perceived accuracy, which shaped their satisfaction with each model. The \textbf{new model} was frequently praised for its precision, as it provided fewer but more relevant results. \textbf{P6} remarked, \textit{"New model gave me much fewer matches again but those matches were much closer in appearance."} Similarly, \textbf{P9} favored the new model, stating, \textit{"I’d prefer the newer one because I am the type of person that prefers quality over quantity."}

However, the smaller pool of results also made inaccuracies more noticeable. \textbf{P8} highlighted this issue, stating, \textit{"The newer model always seemed to have at least 1/4 of the results that were not similar to the test image."} This led some users to prefer the \textbf{old model} for offering a larger pool of choices, despite its occasional inaccuracies. As \textbf{P8} explained, \textit{"The old model is way more preferable... it’s quite a bit more accurate and provides a larger results list."}

In contrast, participants like \textbf{P3} were frustrated with the irrelevant matches that often came with the larger result set from the old model. P3 remarked, \textit{"350 potential matches is too many to scroll through,"} highlighting the tension between having more options and the cognitive load it created.

Ultimately, \textbf{participants who valued precision favored the new model for its focused results}, while those who preferred \textbf{more choices gravitated towards the old model}, even at the cost of sifting through irrelevant options.

\subsubsection{\textbf{Users proposed diverse and inconsistent folk theories about how both models work.}}

\begin{table*}[h]
\centering
\renewcommand{\arraystretch}{1.5}  
\begin{tabularx}{\textwidth}{|p{3cm}|X|X|}
\hline
\textbf{Category} & \textbf{Old Model} & \textbf{New Model} \\ \hline
\textbf{Facial Features} &
“Stronger focus on the features of the face” (P7)\par
“Wide range of characteristics...” (P8)\par
“Results had all types of noses, including skinny ones that did not make sense” (P3)
&
“Better job of finding facial matches with eyes, ears, mouths, etc.” (P5)\par
“Focused on facial hair” (P9)\par
“Focused on eyes and nose shape” (P10)\par
“Focused on hairstyle first (hairstyles can change)” (P3) \\ \hline
\textbf{Occlusion} &
“Low resolution and large facial hair obstructing the face... hampers” (P10)\par
“Obscured by shadow... seemed confused” (P10)\par
“Seemed to focus on the Kepis they wore” (P3)\par
“Puts more emphasis on headwear” (P5)
&
“Low resolution and large facial hair obstructing the face... does not hamper as much” (P10) \\ \hline
\textbf{Latency Perceptions} &
“Felt model was slower” (P3)\par
“Slower than the new model” (P8)\par
“No difference in speed” (P9)
&
“Model is faster as it produces fewer results” (P3)\par
“No difference in speed” (P9) \\ \hline
\textbf{Age Sensitivity} &
“Showed more faces that didn’t look anything like the subject photo; for example, I was shown a photo of a very old man when the subject is obviously young to middle-aged” (P4)
&
“I was amazed to see an older face matching the same face as a young man, as I’ve never seen that happen with the Old Model” (P10)\par
“The new model gave me more results of youthful looking soldiers” (P5) \\ \hline
\textbf{Non-facial Attributes} &
“The model worked well with all image types that I tested (tintype, ambrotype, CDV)” (P8)
&
“Picks up tagged words on the photos, not just the facial characteristics of the sitters” (P4) \\ \hline
\end{tabularx}
\vspace{0.25cm}
\caption{A comparison of user-formed folk theories regarding the old and the new facial recognition models. Quotes from participants illustrate divergent opinions on how each model interprets various aspects such as facial features, occlusions, latency perceptions, age sensitivity, and non-facial attributes.}
\label{tab:folk-theories}
\end{table*}

Participants developed various folk theories to explain how the two models processed different aspects of the images. Their theories diverged significantly, often reflecting personal interpretations of the models' behavior (see Table~\ref{tab:folk-theories}).

\paragraph{\textbf{Facial Feature Recognition}} Participants had differing views on how the models handled facial features. Some, like \textbf{P5} and \textbf{P9}, believed the newer model excelled in matching specific facial details such as \textbf{eyes} and \textbf{ears}. \textbf{P5} noted, \emph{"I feel like the new model does a better job of finding facial matches with eyes, ears, mouths, etc."} Others, such as \textbf{P8}, felt the older model provided a broader analysis, stating, \emph{"I could count on the old model producing results that incorporated a wide range of characteristics (beard style, mustache style, eyes, chin, cheeks, forehead, hairstyle, etc..)."}

\paragraph{\textbf{Occlusion and Headgear}} Theories about \textbf{occlusion} and \textbf{headgear} varied across participants. \textbf{P10} believed the newer model handled occlusions, such as \textbf{shadows} or \textbf{facial hair}, better than the older model: \emph{"Images that are low resolution, poorly lit, or have large facial hair obstructing the face can all hamper this model somewhat, but not nearly as much as the Old Model."} Conversely, \textbf{P5} and \textbf{P3} speculated that the older model placed more emphasis on \textbf{headgear}, with \textbf{P3} noting, \emph{"This model seemed to focus on the Kepis that they wore."}

\paragraph{\textbf{Latency Perceptions}} Although both models had identical retrieval times, participants perceived the older model as slower due to the larger number of results it returned. \textbf{P3} remarked, \emph{"I felt this model was slower as it produced far more results."} In contrast, they found the newer model faster, adding, \emph{"This model is faster as it produces fewer results."} \textbf{P6}, however, observed, \emph{"Both models took about 2 seconds to correctly identify this subject."}

\paragraph{\textbf{Age Sensitivity}} Participants speculated that the models varied in their ability to match faces across different age groups. \textbf{P4} observed that the older model struggled with \textbf{age discrepancies}, stating, \emph{"The old model showed more faces that didn’t look anything like the subject photo; for example, I was shown a photo of a very old man when the subject is obviously young to middle aged."} In contrast, \textbf{P10} praised the newer model for its ability to match faces across different ages: \emph{"I was amazed to see an older face matching the same face as a young man, as I’ve never seen that happen with the Old Model."}

\paragraph{\textbf{Non-Facial Attributes}} Some participants believed the models incorporated \textbf{non-facial attributes} like \textbf{image format} or \textbf{text inscriptions}. \textbf{P8} noted that the older model worked well across various image formats, stating, \emph{"The model worked well with all image types that I tested (tintype, ambrotype, CDV)."} \textbf{P4} speculated that the newer model integrated \textbf{Optical Character Recognition (OCR)}, observing, \emph{"So now I see that the New Model picks up tagged words on the photos, not just the facial characteristics of the sitters."}

\subsection{Summary of Findings}

In the diary study, participants, who had prior experience with the older model on CWPS, were given the option to toggle between the old and new models to compare their experience. The findings revealed that \textbf{while most participants preferred the new model for its higher precision and relevance}, some valued the larger result set of the old model for enabling more exploratory opportunities. Perceived accuracy strongly influenced model preferences, with participants often citing facial feature alignment as a key factor for favoring the new model. However, the \textbf{trade-off between quantity and quality} emerged as a recurring theme, as some participants preferred the broader range of results from the old model despite its occasional inaccuracies. Interestingly, participants \textbf{developed nuanced but inconsistent folk theories about model behavior}, attributing differences in performance to aspects like handling of specific facial features or image quality.

\section{Discussion}

\subsection{Model Updates in Facial Recognition Systems}

Facial recognition systems have found critical applications across diverse domains such as identity verification in gig work~\cite{watkins2023face}, law enforcement~\cite{garvie2022forensic,Harwell2019}, and historical research~\cite{mohanty2019photo,Schwartz2022}, where they support high-stakes decision-making and enhance operational efficiency. However, despite advancements in auditing and model design~\cite{raji2019actionable,raji2022actionable}, major providers like AWS and Azure often release updates without detailing their impact on downstream tasks~\cite{azure,awsrekognition}, leaving end-users to interpret changes independently. Human-AI teaming in facial recognition tasks relies on the complementary strengths of precise algorithmic outputs and human contextual judgment, but prior work and our findings reveal that incomplete user understanding of model behavior and reliance on subjective strategies often limit collaborative performance~\cite{towler2023diverse,howard2020human,carragher2024trust}, highlighting the importance of explicit communication and alignment in such systems.

Across both studies, participants struggled to effectively assess and distinguish between facial recognition models, highlighting challenges in user understanding of model updates. In Study 1, users relied heavily on subjective perceptions of accuracy to form their assessments, which did not translate into performance gains despite the newer model’s improved precision and recall. This reliance on personal experiences persisted in Study 2, where users toggling between explicitly labeled "old" and "new" models developed divergent folk theories about their properties. These findings underscore a critical limitation of silent or minimally communicated updates: without clear guidance, users may fail to align their strategies with model improvements, leaving potential performance benefits unrealized. For instance, while some participants in Study 1 spent additional time deliberating on specific matches with the newer model, this behavioral shift did not consistently enhance task outcomes. For developers, this underscores the necessity of pairing technical advancements with user-centric communication strategies, ensuring updates empower users to fully leverage improved system capabilities.

For high-stakes applications, addressing inconsistent assumptions requires targeted strategies. Educational interfaces~\cite{kusuma2022civil} could address common misunderstandings about model behavior, such as non-determinism and variability. Developers should also prioritize diverse, context-specific datasets (e.g., race, gender, etc.) and move beyond static benchmarking. By combining these datasets with user simulations, streamlined evaluation workflows could produce public reports detailing both model benchmarks and simulated task outcomes, enabling informed adoption and deployment of new models.

\subsection{Communicating Model Updates in AI-Infused Systems}

Communicating model updates is widely recognized as a critical design challenge in AI-infused systems, yet operationalizing this in real-world settings remains underexplored. While human-AI interaction guidelines emphasize transparency in communicating changes~\cite{amershi2019guidelines}, they provide little guidance on what specific information to share --- such as performance metrics, new capabilities, or expected user impacts --- or how to present it effectively. Our studies underscored this challenge, as participants struggled to infer changes between models and showed no significant improvement in task performance. These findings highlight a gap in how updates are communicated, suggesting the need for more granular and targeted strategies that extend beyond simple labels like "new" or "old."

Future work could explore which aspects of model updates most effectively resonate with users. Rather than focusing solely on abstract improvements (e.g., “better accuracy”), updates could be framed in terms of tangible impacts on user experiences, such as faster decision-making or enhanced time spent evaluating AI outputs. Personalized framing strategies, such as categorizing models as "balanced," "creative," or "precise," might also help guide user expectations and foster more intentional interactions, similar to Microsoft’s Copilot modes~\cite{bersano6exploring}. Scenarios where a "new" model performs worse but is framed as an improvement could demonstrate novelty bias~\cite{luo2023catalogue,enwiki:1092714887}, leading users to favor it despite poorer performance. Testing how framing influences trust and skepticism remains critical.

Beyond static communication strategies, dynamic approaches informed by user behavior and perceptions hold significant promise. Vision-language model (VLM)-based agents, leveraging XAI techniques like counterfactual explanations (“what-if” scenarios) and heatmaps, could validate or challenge user-generated folk theories by linking assumptions to the model’s actual logic. For example, counterfactuals could show how alternative queries or results refute or support a folk theory, while heatmaps and concept-based explanations visually map user assumptions to system behavior~\cite{kim2023help}. Large language model (LLM) agents could analyze interaction patterns to identify shallow evaluations or deviations from standard interaction patterns and adaptively intervene with tailored guidance. These dynamic systems could align user strategies with model capabilities, fostering deeper understanding and more effective human-AI collaboration.

\subsection{Generalizing AI Model Update Challenges Across Domains}

Our findings, while situated in the context of facial recognition, reveal broader implications for AI-infused systems where silent or subtle model updates are common. Unlike software updates, which often prompt resistance due to visible disruptions in workflows or undesired changes~\cite{morreale2020my,rula2020s}, AI model updates in black-box systems present unique challenges. Participants in our study struggled to recognize changes, highlighting how opacity exacerbates the difficulty of adapting to improvements. These challenges become even more acute in high-stakes scenarios, where trust, performance, and safety are paramount.

Similar to other algorithm-driven platforms~\cite{register2023attached,milton2023see,devito2018people}, participants crafted multiple, often divergent folk theories to make sense of the system’s behavior (see Table~\ref{tab:folk-theories}). These theories, while sometimes helpful, also led to incorrect assumptions about model behavior, further complicating user adaptation. In high-stakes contexts such as healthcare or autonomous vehicles, reliance on inaccurate folk theories could have severe consequences, from misinterpretation of diagnostic outputs to unsafe decision-making during critical tasks.

For example, in September 2023, Tesla's rollout of an updated Full Self-Driving (FSD) software caused confusion and safety concerns, with drivers reporting unexpected behaviors and unclear changes~\cite{cao_2023}. Similarly, in healthcare, hallucinations in AI transcription tools have introduced inaccuracies in patient records, forcing practitioners to adopt heightened vigilance to avoid potential errors~\cite{edwards2024whisper}. These cases highlight that even silent improvements in AI accuracy may require users to adapt proactively, emphasizing the need for alignment between system behavior and user workflows.

Consistent with prior human-AI teaming work~\cite{chattopadhyay2017evaluating,bansal2019updates}, improved models did not always translate into better outcomes, as users failed to adapt their strategies to take advantage of system improvements. This misalignment underscores a common challenge across domains where users rely on preexisting behaviors rather than adjusting to updated AI capabilities. While prior work~\cite{wang2023watch} demonstrated that users could detect explanation changes during model updates and adjust accordingly, our findings highlight the unique challenges of black-box systems. In the absence of accompanying explanations, users were unable to recognize updates, suggesting that generalizing successful strategies from explainable AI contexts may require significant adaptation for opaque systems. 

In our studies, we observed a contrast between Prolific participants, who focused primarily on AI outputs, and CWPS participants, who integrated contextual reasoning, such as uniforms and locations, alongside AI assistance. Despite their expertise and long-term exposure to the older model, CWPS participants still developed inconsistent folk theories about the updated model’s behavior, illustrating the diverse beliefs users form when interpreting AI systems. For experienced users like those on CWPS, decision-making speed was not a meaningful factor, as their task emphasized thorough analysis over quick judgments. While more work is needed to determine which groups can reliably judge model changes, our findings highlight that good communication is essential for all users, irrespective of expertise, and must be tailored to diverse behaviors and beliefs to better align expectations with AI system updates.

Our work highlights the importance of granular communication in AI-infused systems, particularly for model updates, where simply notifying users of changes is insufficient. While our metrics focus on a visual retrieval task, they underscore broader challenges in AI-assisted decision-making, emphasizing the need to tailor strategies to align user behaviors with evolving system capabilities.

\section{Limitations and Future Work}\label{sec:limitations}

In Study 1, we aimed to evaluate participants' ability to distinguish between models within a controlled environment. However, the study's duration and design may have limited participants' familiarity with the models, as no improvement trends were observed over the seven trials. Prolonged interactions, as seen in real-world applications, could enhance users' ability to detect model changes over time. The design also presented trial-specific details, such as perceived accuracy ratings, which may have encouraged post-hoc justifications rather than reflecting real-time decision-making. A think-aloud study could mitigate this by capturing participants’ reasoning during interactions. While we conducted pre-planned statistical tests for recall, precision, and false positive rates (see Section~\ref{sec:hat_performance}), corrections for multiple comparisons were not applied, though the pre-planned nature and reporting of non-significant results ensured transparency. The study was also not pre-registered, which we acknowledge as a limitation. Furthermore, the lack of pre- and post-study assessments of AI literacy limits our understanding of shifts in participants’ mental models.

The controlled setting may limit ecological validity, as real-world interactions involve prolonged usage, distractions, and higher stakes. Additionally, the participant sample, recruited online, may not represent domain-specific users, such as historians, and the tasks differed from real-world workflows, limiting generalizability. The significant difference in result counts between models, reflecting their native configurations, underscores trade-offs between result quantity and quality. Artificially constraining the older model's results could standardize comparisons but would obscure real-world behaviors. Systems like Civil War Photo Sleuth and Google Search~\cite{Google2024SearchUpdate} illustrate these trade-offs, where updates reshape result characteristics and user experiences. Future work could explore how such variations influence user perceptions and performance. While our study focused on AI-assisted decision-making in a content-based image retrieval (CBIR) setting, we also acknowledge that richer forms of human-AI teaming, such as those involving agentic, iterative, and communicative AI team members~\cite{o2022human}, may lead to different impacts on user dynamics when models are updated.

In Study 2, a deployment study with 10 participants provided insights into model differences but was limited by the small sample size and focus on experienced users, which restricts generalizability. A larger-scale, in-the-wild deployment could yield insights into model adoption patterns and broader behaviors. While the diary study approach captured qualitative feedback, it did not log detailed interaction patterns such as scrolling and toggling. Incorporating interaction logging in future studies could address this gap. Additionally, we did not communicate the benchmarking performance of the models to avoid bias, but future studies could explore how sharing this information impacts user perceptions and performance. Finally, while the study focused on historical photo identification, further research is needed to assess how these findings generalize to other domains or AI-infused systems.

\section{Conclusion}

In this work, we investigated how users distinguish and perceive different black-box AI models, specifically focusing on facial recognition models in the context of historical person identification. We first conducted a benchmarking study and found that the newer, developer-certified model performs better than the older model on a historical dataset. However, despite the discernible differences in capabilities, this did not translate to a significant performance gain in human-AI team performance with the newer model.

In the online experiment we conducted, we observed that crowd workers found it challenging to discern between the models across trials. Notably, they largely relied on their perceived accuracy of the models, instead of focusing on visible model characteristics such as the number of search results and latency. When real-world users of the Civil War Photo Sleuth (CWPS), a facial recognition-based platform for identifying historical photos, were asked to compare the two models, we received mixed responses. While the majority leaned towards the newer model, a subset preferred the older model, and a few expressed a desire to use both models in tandem. These preferences were influenced by various factors including perceived accuracy and cognitive load. Interestingly, users also developed several divergent folk theories surrounding each model's capabilities.

As part of future work, we aim on exploring how individual differences and design interventions for communicating model differences can potentially influence the way people perceive and interact with different models. Overall, our work opens doors for future research in the area of AI-infused systems, offering insights into the potential for smoother model updates and fostering a more harmonious human-AI collaboration.

\begin{acks}
This research was supported by NSF IIS-165196, the Dr. Dennis G. Kafura Graduate Fellowship from the Department of Computer Science at Virginia Tech, and the Center for Human-Computer Interaction at Virginia Tech. We extend our gratitude to all study participants and reviewers who provided valuable feedback during the course of this work. We also thank Dr. Alexandre Filipowicz for early discussions and suggestions, which helped shape the study design and direction.
\end{acks}

\bibliographystyle{ACM-Reference-Format}
\bibliography{sample-base}

\appendix

\section{Appendix}
\subsection{Generative AI Usage}

We used ChatGPT to 1) generate the initial captions and descriptions for the images,  2) for polishing the quality of text, and 3) formatting the tables.

\subsection{Benchmarking the Old and New Facial Recognition Models}\label{appendix:benchmarking}

From Table~\ref{tab:modelbenchmarking}, we find that new model consistently outperforms old model across all evaluated metrics — average precision, recall, and reciprocal rank — on our study dataset. There are several instances where new model was able to retrieve correct matches at higher ranks when old model was unsuccessful. This indicates that new model has a more accurate and refined search capability compared to old model. In addition, new model tends to return a significantly smaller number of results compared to old model, which is indicative of the model's precision.

We also observe that new model consistently retrieves correct matches at higher ranks and with greater confidence scores than old model (see Table~\ref{tab:rankbenchmarking}). This confirms that new model not only identifies matches more accurately but also assigns them higher confidence scores, reflecting its enhanced performance. Importantly, new model also performs better across a diverse range of faces, including those of African American (Photo IDs: 42234, 25008), Hispanic (Photo ID: 41622), and female (Photo IDs: 41631, 41825) subjects. 

\begin{table*}[p]
\footnotesize
\centering
\renewcommand{\arraystretch}{1.5} 
\resizebox{\textwidth}{!}{%
\begin{tabular}{c|cc|cc|cc|cc}
\multirow{2}{*}{\textbf{Photo ID}} &
  \multicolumn{2}{c|}{\textbf{Average Precision}} &
  \multicolumn{2}{c|}{\textbf{Reciprocal Rank}} &
  \multicolumn{2}{c|}{\textbf{Recall}} &
  \multicolumn{2}{c}{\textbf{Total Search Results}} \\ \cline{2-9} 
 &
  \textbf{Old Model} &
  \textbf{New Model} &
  \textbf{Old Model} &
  \textbf{New Model} &
  \textbf{Old Model} &
  \textbf{New Model} &
  \textbf{Old Model} &
  \textbf{New Model} \\ \hline
\textbf{52778} & 0.00          & 0.00          & 0.00          & 0.00          & 0.00          & 0.00          & 625 & 142 \\
\textbf{40914} & \textbf{0.54} & \textbf{1.00} & 1.00          & 1.00          & 0.80          & 0.80          & 785 & 35  \\
\textbf{51831} & \textbf{0.00} & \textbf{1.00} & 0.00          & 1.00          & \textbf{0.00} & \textbf{1.00} & 775 & 16  \\
\textbf{29626} & \textbf{0.35} & \textbf{1.00} & 1.00          & 1.00          & 0.75          & 0.75          & 789 & 17  \\
\textbf{41787} & \textbf{0.63} & \textbf{1.00} & 1.00          & 1.00          & 1.00          & 1.00          & 792 & 17  \\
\textbf{30357} & 0.00          & 1.00          & \textbf{0.00} & \textbf{1.00} & \textbf{0.00} & \textbf{1.00} & 763 & 12  \\
\textbf{34096} & 1.00          & 1.00          & 1.00          & 1.00          & 1.00          & 1.00          & 25  & 11  \\
\textbf{19561} & 1.00          & 1.00          & 1.00          & 1.00          & 1.00          & 1.00          & 846 & 15  \\
\textbf{46204} & 1.00          & 1.00          & 1.00          & 1.00          & 1.00          & 1.00          & 14  & 11  \\
\textbf{19595} & 0.00          & 0.00          & 0.00          & 0.00          & 0.00          & 0.00          & 144 & 15  \\
\textbf{29405} & \textbf{0.02} & \textbf{0.75} & \textbf{0.04} & \textbf{1.00} & 0.71          & 0.71          & 849 & 32  \\
\textbf{46459} & 0.00          & 1.00          & \textbf{0.00} & \textbf{1.00} & \textbf{0.00} & \textbf{1.00} & 24  & 13  \\
\textbf{42234} & \textbf{0.05} & \textbf{1.00} & \textbf{0.05} & \textbf{1.00} & 1.00          & 1.00          & 108 & 10  \\
\textbf{24322} & 1.00          & 1.00          & 1.00          & 1.00          & 1.00          & 1.00          & 26  & 5   \\
\textbf{25008} & \textbf{0.75} & \textbf{1.00} & 1.00          & 1.00          & 1.00          & 1.00          & 63  & 6   \\
\textbf{24356} & 0.00          & 0.00          & 0.00          & 0.00          & 0.00          & 0.00          & 38  & 11  \\
\textbf{41901} & 0.00          & 0.00          & 0.00          & 0.00          & 0.00          & 0.00          & 37  & 9   \\
\textbf{41622} &
  \textbf{0.00} &
  \textbf{1.00} &
  \textbf{0.00} &
  \textbf{1.00} &
  \textbf{0.00} &
  \textbf{1.00} &
  695 &
  11 \\
\textbf{41631} & 1.00          & 1.00          & 1.00          & 1.00          & \textbf{0.50} & \textbf{1.00} & 4   & 7   \\
\textbf{41825} & \textbf{0.92} & \textbf{1.00} & 1.00          & 1.00          & 1.00          & 1.00          & 156 & 7   \\ \hline
\textbf{Mean} &
  \textbf{0.41} &
  \textbf{0.79} &
  \textbf{0.50} &
  \textbf{0.80} &
  \textbf{0.54} &
  \textbf{0.76} &
  \textbf{377.90} &
  \textbf{20.10}
\end{tabular}%
}
\vspace{10pt} 
\caption{Model Benchmarking Results: Average  precision, reciprocal rank, recall, and the total number of search results for all photos in the study dataset.}
\label{tab:modelbenchmarking}
\end{table*}

\begin{table*}[h]
\centering
\renewcommand{\arraystretch}{2} 
\resizebox{\textwidth}{!}{%
\begin{tabular}{c|cc|cc}
\multirow{2}{*}{\textbf{Photo ID}} &
  \multicolumn{2}{c|}{\textbf{Ranks of Correct Matches}} &
  \multicolumn{2}{c}{\textbf{Confidence Scores of Correct Matches}} \\ \cline{2-5} 
      & \textbf{Old Model}   & \textbf{New Model} & \textbf{Old Model}                         & \textbf{New Model}                         \\ \hline
\textbf{52778} & {[}{]}             & {[}{]}           & {[}{]}                                   & {[}{]}                                   \\
\textbf{40914} & {[}1, 2, 33, 68{]} & {[}1, 2, 3, 4{]} & {[}80.11\%, 78.98\%, 68.78\%, 67.32\%{]} & {[}87.85\%, 84.79\%, 77.08\%, 76.39\%{]} \\
\textbf{51831} & {[}{]}             & {[}1{]}          & {[}{]}                                   & {[}87.66\%{]}                            \\
\textbf{29626} & {[}1, 92, 187{]}   & {[}1, 2, 3{]}    & {[}73.86\%, 59.54\%, 57.98\%{]}          & {[}77.74\%, 75.88\%, 75.53\%{]}          \\
\textbf{41787} & {[}1, 2, 6, 107{]} & {[}1, 2, 3, 4{]} & {[}74.34\%, 72.85\%, 66.06\%, 58.58\%{]} & {[}89.26\%, 81.85\%, 79.86\%, 79.86\%{]} \\
\textbf{30357} & {[}{]}             & {[}1{]}          & {[}{]}                                   & {[}74.61\%{]}                            \\
\textbf{34096} & {[}1{]}            & {[}1{]}          & {[}63.01\%{]}                            & {[}91.24\%{]}                            \\
\textbf{19561} & {[}1{]}            & {[}1{]}          & {[}77.25\%{]}                            & {[}90.22\%{]}                            \\
\textbf{46204} & {[}1{]}            & {[}1{]}          & {[}59.25\%{]}                            & {[}91.27\%{]}                            \\
\textbf{19595} & {[}{]}             & {[}{]}           & {[}{]}                                   & {[}{]}                                   \\
\textbf{29405} &
  {[}23, 87, 149, 179, 791{]} &
  {[}1, 2, 3, 8, 20{]} &
  {[}61.34\%, 58.43\%, 56.72\%, 56.09\%, 51.08\%{]} &
  {[}81.44\%, 79.48\%, 71.06\%, 64.86\%, 52.23\%{]} \\
\textbf{46459} & {[}{]}             & {[}1{]}          & {[}{]}                                   & {[}75.50\%{]}                            \\
\textbf{42234} & {[}19{]}           & {[}1{]}          & {[}54.78\%{]}                            & {[}83.86\%{]}                            \\
\textbf{24322} & {[}1{]}            & {[}1{]}          & {[}78.66\%{]}                            & {[}93.84\%{]}                            \\
\textbf{25008} & {[}1, 4{]}         & {[}1, 2{]}       & {[}70.51\%, 59.59\%{]}                   & {[}89.62\%, 76.02\%{]}                   \\
\textbf{24356} & {[}{]}             & {[}{]}           & {[}{]}                                   & {[}{]}                                   \\
\textbf{41901} & {[}{]}             & {[}{]}           & {[}{]}                                   & {[}{]}                                   \\
\textbf{41622} & {[}{]}             & {[}1{]}          & {[}{]}                                   & {[}75.02\%{]}                            \\
\textbf{41631} &
  {[}1, 2, 3{]} &
  {[}1, 2, 3, 4, 5, 6{]} &
  {[}69.19\%, 68.91\%, 59.14\%{]} &
  {[}92.18\%, 92.11\%, 90.06\%, 78.26\%, 78.01\%, 75.02\%{]} \\
\textbf{41825} & {[}1, 2, 4{]}      & {[}1, 2, 3{]}    & {[}70.10\%, 64.84\%, 61.56\%{]}          & {[}80.13\%, 79.60\%, 75.09\%{]}         
\end{tabular}%
}
\vspace{10pt} 
\caption{Model Benchmarking: Ranks and confidence scores of correct matches retrieved for all photos in the dataset. }
\label{tab:rankbenchmarking}
\end{table*}

\end{document}